\documentclass[preprint,12pt]{elsarticle}

\usepackage{amsmath,amssymb}
\usepackage{graphicx}
\usepackage{booktabs}
\usepackage{lineno}
\usepackage{xcolor}

\newcommand{\dd}{\mathrm{d}}
\newcommand{\ii}{\mathrm{i}}
\newcommand{\ee}{\mathrm{e}}
\journal{Computer Methods and Programs in Biomedicine}

\begin{document}

\begin{frontmatter}

\title{An interpretable model of spectral scattering of arterial pulse waves in the circle of Willis
encodes occlusion location}

\author[pku]{Xun Huang\corref{cor1}}
\cortext[cor1]{Corresponding author. }
\ead{huangxun@pku.edu.cn}
\affiliation[pku]{organization={State Key Laboratory of Turbulence and
Complex Systems, School of Mechanics and Engineering Science, Peking
University}, city={Beijing}, country={China}}

\begin{abstract}
\textit{Background and objective.} Carotid Doppler ultrasound is the most
widely available bedside probe of cerebral haemodynamics, and
machine-learning classifiers fed carotid velocity spectra can
localise intracranial occlusions---but at the price of black-box models,
thousands of training samples, and fragility to anatomical variants. We
ask whether the same information can be obtained from an interpretable
physical model. \textit{Methods.} We formulate a linear frequency-domain
one-dimensional model of a 26-segment circle-of-Willis network in which
each cardiac harmonic propagates on a Womersley transmission line and is
scattered independently by a lesion. The model is validated against a
nonlinear one-dimensional solver and three-dimensional stenosis
computations at two severities, coupled to a physics-based
synthetic spectral-Doppler pipeline, and used to generate a
9{,}600-case virtual cohort for
occlusion-localisation classification with physics-guided features.
\textit{Results.}
Mean flow divisions agree with the nonlinear solver to three significant
figures, harmonic magnitudes within a ratio of 0.90--1.04, and carotid
waveforms to 6.1\,\% in relative $L_2$ norm after a single kinematic
area correction. The benchmark
reproduces healthy flow splits within 1--9\,\%, collateral-channel flows
within 11--16\,\%, and the collateral-pathway hierarchy. Physics-guided
harmonic-ratio features localise twelve lesion classes at 95.1\,\%
accuracy from five training samples per class---over an order of
magnitude fewer than waveform-driven convoluted neural networks; augmentation partially
restores robustness to measurement noise and anatomical variants.
\textit{Conclusions.} The framework explains the physical origin of the
empirically discriminative 2--12\,Hz Doppler band and makes occlusion
localisation interpretable and sample-efficient. 
%prospective validation on patient spectra is the decisive next test.
\end{abstract}


\end{frontmatter}

%\linenumbers

%=====================================================================
\section{Introduction}
\label{sec:intro}

%\subsection{Clinical motivation}

Stroke remains one of the leading causes of death and long-term disability
worldwide, and health-system readiness for its acute management is an
explicit priority of current global health policy \citep{who2026}. For
acute ischaemic stroke caused by a large-vessel occlusion, the single most
valuable piece of pre-hospital information is the {location} of the
occlusion, because it determines triage to a thrombectomy-capable centre.
Yet the only bedside instrument that samples cerebral haemodynamics
directly---the carotid Doppler examination---returns
velocity spectra whose connection to the occlusion site has so far been
established only empirically. To address this critical gap, the present study establishes an analytical model and provides corresponding simulation-based validation. 

Carotid duplex scanning is cheap, portable
and near-universal: scoping reviews of scanning protocols document its
routine use worldwide, while deploring the lack of
standardised diagnostic criteria beyond proximal stenosis grading
\citep{mukabagorora2025}. The clinical Doppler literature records that
proximal occlusions rewrite the carotid waveform in recognisable
ways---side-asymmetric peak velocities in innominate-artery disease
\citep{grant2006}, the tardus--parvus morphology (a spectral Doppler waveform characterised by a prolonged systolic acceleration time and a reduced peak systolic velocity with diminished pulsatility) betraying a proximal
obstruction at a site without local stenosis \citep{park2019}, and the
spectral-broadening criteria by which stenoses are graded
\citep{nuffer2017}---but no quantitative, mechanics-based link between an
{intracranial} lesion and the {extracranial} spectrum is part of
current practice.

%\subsection{1D pulse-wave modelling and diagnosis}

One‑dimensional models of pulse‑wave propagation in compliant arterial networks constitute a mature forward‑modelling technology \citep{sherwin2003pulse, fqv2009, mynard2015}. The circle of Willis ---a ring of arteries that connects the major cerebral circulations---has been a canonical application since the nonlinear network study by Alastruey et al. \citep{alastruey2007}, which quantified how anatomical variants and occlusions redistribute cerebral blood flow. 
The branched transmission-line structure itself has been analysed in depth
\citep{fullana2009}. 

One-dimensional models have been used to attribute aortic-root reflections
to upper- and lower-body vessels \citep{abdullateef2023}, to estimate
central pulse-wave velocity from radial pulse-wave analysis
\citep{yao2022}, and---most directly relevant to the present work---to
diagnose common-carotid stenosis from mechanical waves scattered by the
lesion \citep{floresgeronimo2022}. On the measurement side, ultrasound has
been used to estimate central waveforms non-invasively
\citep{zhou2024}; Doppler-derived inflow conditions have been shown to
matter for downstream haemodynamic metrics \citep{rutten2026}; and
personalised zero-dimensional carotid models have been fitted to
individuals \citep{khan2021}. What remains missing is the {inverse}
reading of the same physics for the intracranial network: a closed-form
statement of how a localised lesion of the circle of Willis rewrites the pulse-wave
spectrum at an accessible carotid measurement station.

%\subsection{Virtual cohorts and waveform classification}

A complementary, data-driven research approach generates virtual patient
cohorts from haemodynamic simulators and trains classifiers on the
synthetic waveforms. The methodological precedent is the study of
{\c{S}}en et al.\ \citep{sen2024}, who trained convolutional neural
networks on synthetic common-carotid velocity waveforms generated by a
nonlinear one-dimensional solver, reaching region-detection true-positive
rates above 95\,\% noise-free and about 80\,\% under 20\,\% measurement
noise. Three of their findings bound the present problem: (1) classifiers fed
with carotid {velocity} waveforms reach $F_1\approx0.95$ for
large-vessel-occlusion detection whereas pressure waveforms remain below
$0.5$, singling out the velocity spectrum as the informative channel; (2) the
discriminative content concentrates in the 2--12\,Hz band; and (3) performance
saturates only at roughly 7{,}000 training samples and collapses on
realistic asymmetric geometries for medium- and small-vessel occlusions.

More broadly, the virtual‑cohort methodology has become an active area of research, spanning applications from stochastic virtual patients with temporal evolution for ventilation trials \citep{ang2023}, to in‑silico cohorts for virtual coronary revascularisation \citep{lyu2025}, and to realistically generated synthetic cerebrovascular trees for perfusion simulation \citep{rundfeldt2024}. Waveform
classification by machine learning is likewise established---peripheral
pulse waves predict cardiac function \citep{wang2022}, deep learning
delivers fast surrogate haemodynamic indices \citep{liu2024ifr}, and
photoplethysmogram features classify blood-pressure states
\citep{mejiamejia2021}---but the features are learned, not derived, and
the sample complexity and domain fragility documented by
{\c{S}}en et al.\ underscore the necessity of additional studies. Cerebral-flow modelling specific to
the circle of Willis has quantified the haemodynamic consequences of
internal-carotid-artery stenosis \citep{sun2024cow} and delivered
real-time model-based perfusion estimates for ischaemic stroke
\citep{sun2024perfusion}, again in the forward direction.

%\subsection{The present contribution}

This paper connects the two threads. We utilise the classical theoretical method in 
fluid mechanics and transfer the
spectral-decomposition scaffolding of Bell's theory of lee-wave generation
by simple-harmonic flow over topography \citep{bell1975} to the compliant
circle of Willis network. More specifically, linearisation about the periodic cardiac state decouples the
cardiac harmonics exactly, so each harmonic propagates on a Womersley
transmission line \citep{womersley1955} and is scattered independently by
a lesion, modelled as a branch closure (occlusion) or as a linearised
Young--Tsai series impedance (stenosis) \citep{young1973}. The framework
yields closed-form, falsifiable signatures---a spectral ripple whose
spacing $\Delta f=c_{\mathrm{eff}}/(2L_{\mathrm{eff}})$ encodes the
occluder distance, and harmonic-amplitude-ratio redistributions that
encode site and severity---and it explains why the empirically
discriminative 2--12\,Hz band is where it is.

The remaining part of the paper is organised accordingly. Section~\ref{sec:methods} develops the
model and the lesion representations. Section~\ref{sec:validation} gives the validation design details
 and the virtual-cohort classification experimental set-ups. 
Section~\ref{sec:results} reports the associated results of the validation, the external benchmark,
the spectral signatures and synthetic spectrograms, and the classifier
performance. Section~\ref{sec:discussion} discusses the clinical and
methodological implications and the limitations, and
Section~\ref{sec:conclusions} concludes the whole work.

\begin{figure}[t]
\centering
\includegraphics[width=\textwidth]{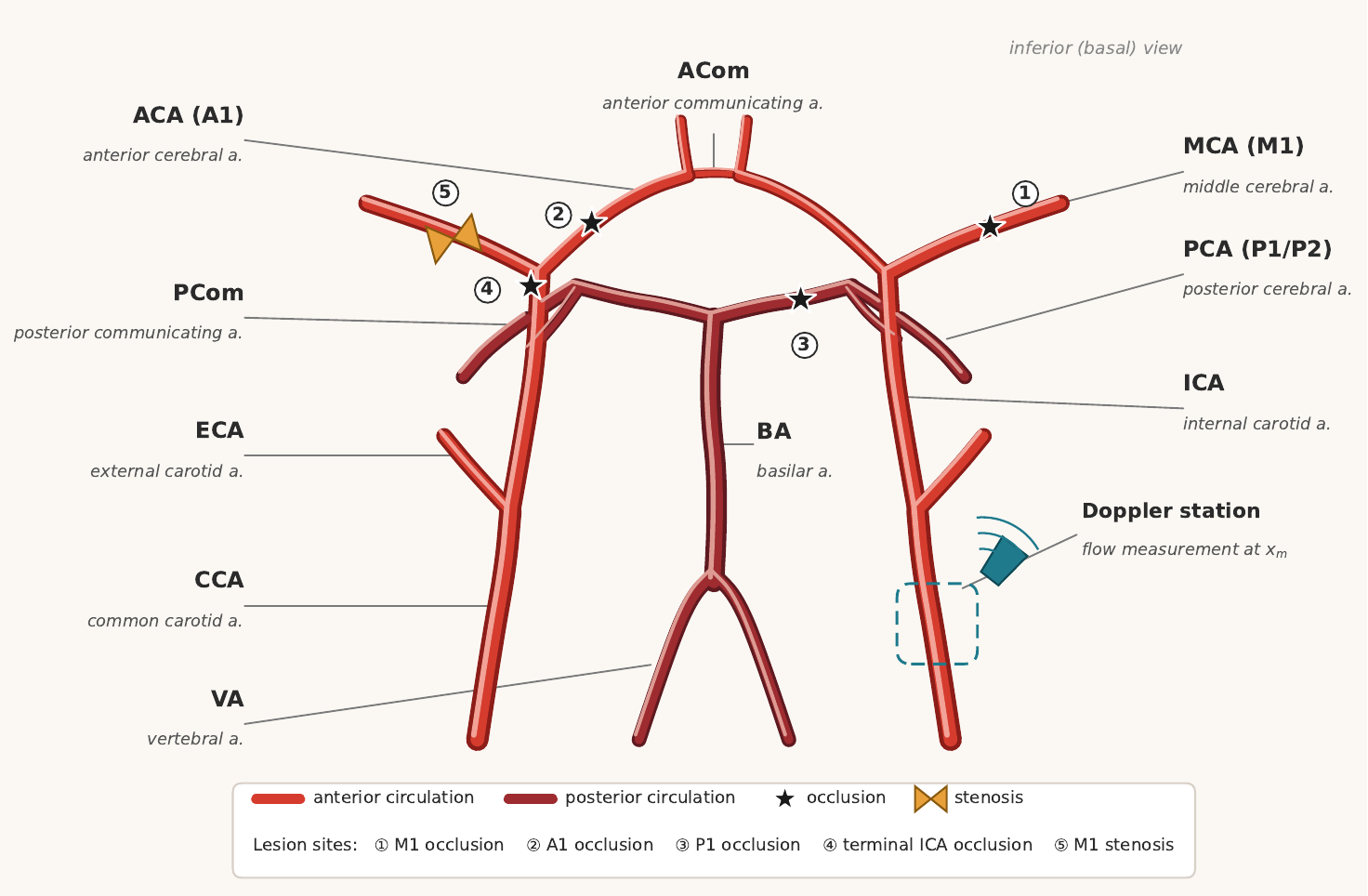}
\caption{Model network and lesion sites. The reduced circle-of-Willis
network comprises four prescribed-flow inlets (both common carotid
arteries, CCA, and both vertebral arteries, VA), compliant segments joined
at bifurcation nodes, and terminal three-element Windkessel loads. Filled
markers indicate the five lesion classes of Table~\ref{tab:scen}
(bilateral pairs share one marker); the Doppler measurement stations sit
at the CCA midpoints. ECA/ICA: external/internal carotid artery; M1/A1/P1:
first segments of the middle/anterior/posterior cerebral arteries; ACoA,
PCoA: anterior/posterior communicating arteries; BA: basilar artery.}
\label{fig:schematic}
\end{figure}
%
%=====================================================================
\section{Methods}
\label{sec:methods}

The model network (Fig.~\ref{fig:schematic}) retains the twenty-six
compliant segments of the reduced circle of Willis network of Alastruey et al.\
\citep{alastruey2007}: per side, a common carotid artery (CCA) feeding an
external carotid branch and a two-segment internal carotid artery (ICA),
the posterior communicating artery (PCoA) attaching at the ICA junction,
the first middle- and anterior-cerebral segments (M1, A1) at the ICA
terminus, the anterior communicating artery (ACoA) between the A1/A2
junctions, and the two vertebral arteries (VA) joining into the basilar
artery (BA) and the first posterior-cerebral segments (P1).

\begin{figure}[t]
\centering
\includegraphics[width=\textwidth]{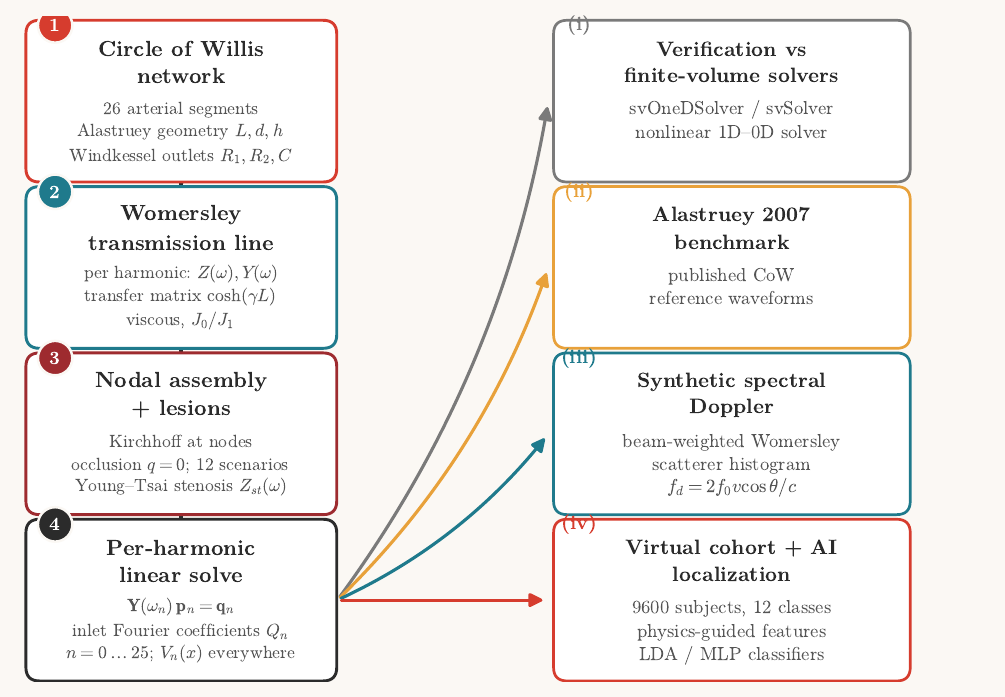}
\caption{Method pipeline. Left to right: the 26-segment
circle-of-Willis network is formulated as a Womersley transmission-line
model, assembled with bifurcation nodes, terminal Windkessel loads and
lesion representations (branch closure for occlusions, linearised
Young--Tsai series impedance for stenoses), and solved harmonic by
harmonic. %The solved field feeds four applications: validation against
%svOneDSolver and a three-dimensional svSolver stenosis computation, the
%external benchmark against Alastruey et al.\ \citep{alastruey2007},
%synthetic spectral-Doppler rendering, and virtual-cohort classification
%with physics-guided features.}
}
\label{fig:pipeline}
\end{figure}
The full method chain is summarised in Fig.~\ref{fig:pipeline}. A
twenty-six-segment circle-of-Willis network with measured-type inflow
waveforms and Windkessel terminal loads is linearised about the periodic
cardiac state, so that each cardiac harmonic propagates independently on a
Womersley transmission line; bifurcation nodes, terminal impedances and
lesion representations are assembled into one sparse linear system per
harmonic, whose solution gives pressure and flow spectra at every node.
Four application branches consume the same solved field: validation
against independent nonlinear one-dimensional and three-dimensional
solvers, an external benchmark against a published circle-of-Willis study,
a physics-based synthetic spectral-Doppler pipeline, and a virtual-cohort
occlusion-localisation classifier built on physics-guided features.

\subsection{Linear frequency-domain network model}
\label{sec:model}

%\subsubsection{Network, parameters and boundary conditions}

A succinct version of this subsection has been submitted elsewhere as a Letter and is included here for the completeness of the present paper. All geometric, elastic and terminal-load parameters are from the literature
\citep{alastruey2007}, and 
wall thicknesses follow the regression of \citep{alastruey2007}. The
second middle-cerebral segment (M2) is not part of the original network
and is retained as an approximation ($L=0.05$\,m, $d=2.4$\,mm) so that the
middle-cerebral Windkessel load keeps its verified values. Four inlets
prescribe measured-type flow waveforms at the two CCA roots (mean
5.0\,mL\,s$^{-1}$ each) and the two VA roots (1.0\,mL\,s$^{-1}$ each), the
reference heart rate is 60\,bpm, and each terminal microvascular bed is
closed by a three-element Windkessel load. The resulting carotid reference
wave speeds are 5.5--6.5\,m\,s$^{-1}$. The Doppler measurement station is
the midpoint of each CCA segment, the standard carotid examination
location. The healthy network reproduces the ordering and approximate
magnitudes of in-vivo regional flow measurements \citep{zarrinkoob2015}.

%\subsubsection{Governing equations and linearisation}

For each vessel segment with axial coordinate $x$ and time $t$, the
one-dimensional mass and momentum balances for the cross-sectional area
$A$, volume flow rate $Q$ and mean pressure $P$ read
\citep{alastruey2007, fqv2009}
\begin{equation}
\frac{\partial A}{\partial t}+\frac{\partial Q}{\partial x}=0,
\qquad
\frac{\partial Q}{\partial t}
+\frac{\partial}{\partial x}\!\left(\frac{Q^{2}}{A}\right)
=-\frac{A}{\rho}\frac{\partial P}{\partial x}
-\frac{8\pi\mu}{\rho}\frac{Q}{A},
\label{eq:massmom}
\end{equation}
with blood density $\rho$, viscosity $\mu$ and a blunt velocity profile
(momentum correction factor one). The tube law is the linear-elastic
relation $P-P_{\mathrm{ext}}=(\beta/A_0)(\sqrt{A}-\sqrt{A_0})$ with
$\beta=\sqrt{\pi}hE/(1-\nu_p^2)$, giving the reference wave speed
\begin{equation}
c_{0}^{2}=\frac{A_{0}}{\rho}\,
\frac{\partial P}{\partial A}\bigg|_{A_{0}}
=\frac{Eh}{2\rho r_{0}(1-\nu_p^{2})}.
\label{eq:mk}
\end{equation}
Writing each field as a periodic mean plus a perturbation and retaining
first order in the perturbations is justified by the parameter regime:
carotid-to-cerebral values give $U/c\approx0.05$--$0.15$ and radial
strains below ten per cent, so the convective and area-coupling terms are
of relative order $U/c$ and drop out. The perturbation system is
\begin{equation}
\frac{\partial a}{\partial t}+\frac{\partial q}{\partial x}=0,
\qquad
\frac{\rho}{A_{0}}\frac{\partial q}{\partial t}
+\frac{\partial p}{\partial x}
=-\frac{8\pi\mu}{A_{0}^{2}}\,q,
\qquad
p=\frac{\rho c_{0}^{2}}{A_{0}}\,a .
\label{eq:lin}
\end{equation}
Because the coefficients are time-independent, the Fourier expansion
$q(x,t)=\sum_n q_n(x)\,\ee^{\ii n\omega_h t}$ (with $\omega_h=2\pi/T$ the
cardiac angular frequency) yields one autonomous problem per harmonic:
\emph{the cardiac harmonics decouple exactly at linear order}. This is the
mathematical sense in which the arterial problem is simpler than Bell's
geophysical original \citep{bell1975}: the oscillatory advection that
couples harmonics there is absorbed by the linearisation here.

%\subsubsection{Womersley transmission line and network assembly}

For harmonic $n$ at $\omega=n\omega_h$, (\ref{eq:lin}) takes the
transmission-line form
\begin{equation}
-\frac{\dd p_n}{\dd x}=Z(\omega)\,q_n,
\qquad
-\frac{\dd q_n}{\dd x}=Y(\omega)\,p_n,
\label{eq:telegraph}
\end{equation}
whose series impedance follows from the radially resolved oscillatory
flow, i.e.\ Womersley's no-slip solution of the axial momentum balance in
a straight circular tube \citep{womersley1955}:
\begin{equation}
Z(\omega)=\frac{\ii\omega\rho}{A_{0}}\,\bigl[1-F_{J}(\alpha_W)\bigr]^{-1},
\qquad
F_{J}(s)=\frac{2J_{1}(s)}{sJ_{0}(s)},
\label{eq:FJ}
\end{equation}
with $J_0,J_1$ Bessel functions of the first kind of order 0 and 1, respectively,
$s=\ii^{3/2}\alpha_W$, and $\alpha_W=r_{0}\sqrt{\omega\rho/\mu}$ the
Womersley number. The low-frequency limit $Z\to8\pi\mu/A_0^2$ recovers the
Poiseuille resistance, so the steady friction of (\ref{eq:lin}) is
recovered smoothly; the shunt admittance is $Y(\omega)=\ii\omega C_w$ with
wall compliance per unit length $C_w=A_0/(\rho c_0^2)$. The propagation
constant and characteristic impedance
\begin{equation}
\gamma(\omega)=\sqrt{Z(\omega)Y(\omega)}=\alpha(\omega)+\ii k(\omega),
\qquad
Z_{c}(\omega)=\sqrt{Z(\omega)/Y(\omega)},
\label{eq:gamma}
\end{equation}
define the attenuation and wavenumber: the Womersley dispersion raises the
phase speed $\omega/k$ monotonically from $c_0$ while the attenuation
grows as $\alpha\propto\sqrt{\omega}$---all frequencies propagate, high
frequencies dissipate strongly, which already anticipates that
lesion-generated information must live in the low harmonics. A uniform
segment of length $L$ maps the state $[p;q]$ through the transfer matrix
\begin{equation}
T(\omega)=
\begin{bmatrix}
\cosh\gamma L & Z_c\sinh\gamma L\\
Z_c^{-1}\sinh\gamma L & \cosh\gamma L
\end{bmatrix},
\label{eq:T}
\end{equation}
bifurcation nodes enforce pressure continuity and flow conservation, and
terminal beds contribute the Windkessel impedance
$Z_W(\omega)=R_1+R_2/(1+\ii\omega R_2C)$, whose elements carry direct
physiological meaning: $R_1$ matches the terminal segment's characteristic
impedance, $R_2$ sets the territory's mean pressure--flow relation, and
$C$ buffers pulsatile inflow into near-steady perfusion. Assembling
(\ref{eq:T}) per harmonic gives one sparse linear system per $n$, whose
solution yields $(p_n,q_n)$ at every node; the time-domain field follows
by Fourier synthesis. %A per-harmonic solve costs milliseconds on one CPU core.

\subsection{Transfer function \& spectral ripple}
\label{sec:ripple}

Let the measurement station $x_m$ lie upstream of the occlusion site,
joined to it by an effective round-trip path of length $L_{\mathrm{eff}}$.
Decomposing the harmonic field into the wave incident from the heart side
and the wave returned by the occluded subtree gives, under the single
dominant-path approximation,
\begin{equation}
\hat V_n(x_m)=\hat V_n^{\mathrm{inc}}(x_m)\Bigl[1+\Gamma_{\mathrm{eff}}(n\omega_h)\,
\ee^{-2\gamma(n\omega_h)L_{\mathrm{eff}}}\Bigr]+\text{remaining paths},
\label{eq:tf}
\end{equation}
where $\Gamma_{\mathrm{eff}}$ is the effective reflection coefficient of
the occluded subtree. Expanding the squared modulus term by term with
$\gamma=\alpha+\ii k$,
\begin{equation}
|\hat V_n|^{2}\approx|\hat V_n^{\mathrm{inc}}|^{2}
\Bigl[1+|\Gamma_{\mathrm{eff}}|^{2}\ee^{-4\alpha L_{\mathrm{eff}}}
+2|\Gamma_{\mathrm{eff}}|\ee^{-2\alpha L_{\mathrm{eff}}}
\cos\bigl(2k L_{\mathrm{eff}}+\phi_\Gamma\bigr)\Bigr].
\label{eq:ripple}
\end{equation}
Equation~(\ref{eq:ripple}) is the structural core of the framework: the
interference term is periodic in frequency, so \emph{the occlusion site is
encoded as a spectral-ripple spacing}
\begin{equation}
\Delta f=\frac{c_{\mathrm{eff}}}{2L_{\mathrm{eff}}},
\qquad c_{\mathrm{eff}}(\omega)=\omega/k(\omega),
\label{eq:deltaf}
\end{equation}
while \emph{the occlusion severity is encoded as the ripple depth}
$2|\Gamma_{\mathrm{eff}}|\ee^{-2\alpha L_{\mathrm{eff}}}$ and in a
redistribution of harmonic amplitude ratios. When several reflection paths
contribute, (\ref{eq:ripple}) becomes a sum of cosines whose spacings are
separable by a cepstrum of $\log|\hat V(f)|$, each peak marking a
round-trip time $2L_i/c_{\mathrm{eff}}$.

\subsection{Lesion modelling: occlusions and stenoses as scattering sites}
\label{sec:lesions}

Pathology enters the linear network in exactly two ways
(Table~\ref{tab:scen}). A complete occlusion replaces the proximal node
condition of the blocked branch by $q_n=0$ for all $n$: to the incident
pressure wave the branch is an open termination with reflection
coefficient tending to $+1$, and the occlusion acts as a newly created
strong reflector that rewrites the harmonic amplitude and phase
distribution of the whole network. For the carotid-terminus scenario the
distal ICA is retained as a closed dead-end stub (shunt admittance
$\tanh(\gamma L)/Z_c$ at its proximal node), so that the stub's
standing-wave load is represented exactly. A partial stenosis is
represented by the Young--Tsai lumped pressure drop \citep{young1973},
standard in one-dimensional cerebral models \citep{mynard2015},
\begin{equation}
\Delta P
=
\frac{K_t\mu}{D_{0}}\,u
+\frac{\rho K_{v}}{2}\Bigl(\frac{A_{0}}{A_{1}}-1\Bigr)^{2}u\,|u|
+\rho K_{u}L_{s}\frac{\partial u}{\partial t},
\qquad u=\frac{Q}{A_{0}},
\label{eq:yt}
\end{equation}
whose three terms are, in order, the quasi-steady viscous loss along the
lesion, the inertial (Bernoulli-type) loss of the separated
post-stenotic jet---which dominates for haemodynamically significant
stenoses---and the inertance of the blood column within the lesion;
$K_t,K_v,K_u$ are tabulated in \citep{young1973}. Linearising
(\ref{eq:yt}) about the mean flow $\bar Q$ yields the series impedance
\begin{equation}
Z_s(\omega)=\underbrace{\frac{K_t\mu}{D_0 A_0}
+\rho K_{v}\Bigl(\frac{A_0}{A_1}-1\Bigr)^{2}\frac{\bar Q}{A_0^{2}}}
_{\text{effective steady resistance }R_s}
\;+\;\ii\omega\,\rho K_{u}\frac{L_s}{A_0},
\label{eq:Zs}
\end{equation}
inserted at the lesion location. The standard lumped scattering
coefficients of a series impedance on a line,
\begin{equation}
\Gamma_s=\frac{Z_s}{Z_s+2Z_c},
\qquad
\mathcal{T}_s=\frac{2Z_c}{Z_s+2Z_c},
\label{eq:gammas}
\end{equation}
then apply, with $|\Gamma_s|\to1$ as the area stenosis degree
$s_A=1-A_1/A_0$ grows, connecting smoothly to the complete-occlusion
limit; the discarded quadratic term couples harmonic $n$ to harmonics
$n\pm m$ at second order and is the leading source of linearisation
error quantified in Section~\ref{sec:resval}.

\begin{table}[t]
\centering
\caption{Twelve lesion scenarios. An occlusion sets the perturbation flow
to zero at the proximal node of the named branch; a stenosis inserts the
series impedance (\ref{eq:Zs}). M1/A1/P1: first segments of the middle,
anterior and posterior cerebral arteries; ICA terminus: distal end of the
internal carotid artery.}
\label{tab:scen}
\footnotesize
\begin{tabular}{lll}
\toprule
Scenario & Site & Implementation \\
\midrule
healthy & --- & baseline network \\
occlusion, L/R M1 & M1 segment & $q_n=0$ at M1 origin \\
occlusion, L/R A1 & A1 segment & $q_n=0$ at A1 origin \\
occlusion, L/R P1 & P1 segment & $q_n=0$ at P1 origin \\
occlusion, L/R ICA terminus & distal ICA & $q_n=0$ at terminus (stub) \\
stenosis, L M1, $s_A=50/75/90\%$ & M1 segment & $Z_s(\omega)$ of (\ref{eq:Zs}) \\
\bottomrule
\end{tabular}
\end{table}

\section{Validation Designs}
\label{sec:validation}

The linear solver was verified analytically and then compared with two
independent nonlinear solvers of the SimVascular open-source pipeline
\citep{updegrove2017}, and with the published circle-of-Willis benchmark
of Alastruey et al.\ \citep{alastruey2007}.

\subsection{Analytic checks}
\label{sec:analytic}

First, unit tests immediately confirm: (i) the Poiseuille low-frequency limit of equation~(\ref{eq:FJ}) to a relative error of $10^{-8}$; (ii) the standing-wave
resonances of a single closed tube, whose positions match
$\mathrm{Im}(\gamma)L=k\pi$ within 1.5\,\% over seven resonances (measured
spacing 2.57\,Hz against $c_{\mathrm{eff}}/2L=2.46$\,Hz); (iii)
$\det T=1$ to $3\times10^{-16}$ and symmetry of the two-port admittance;
(iv) transfer-impedance reciprocity of the assembled network; and (v)
passivity ($\mathrm{Re}\,\gamma>0$) and the decay rate of a matched tube.

\subsection{Nonlinear one-dimensional solver}
\label{sec:sv1d}

svOneDSolver integrates equation~(\ref{eq:massmom}) without linearisation using a
space-time finite-element method---piecewise linear in space,
discontinuous Galerkin in time, with Galerkin/least-squares
stabilisation and a modified Newton--Raphson solve \citep{wan2002,
pfaller2022}. The deck reproduces the network of
Section~\ref{sec:model} exactly: the same twenty-six segments, four
inlets driven by the same 201-point periodic waveforms, and the same
Windkessel loads; each segment carries an Olufsen-type wall law whose
stiffness is chosen so that the reference wave speed matches equation~(\ref{eq:mk}) exactly. The discretisation uses 1\,mm elements, a 1\,ms
time step and a Newton tolerance of $10^{-8}$; each run covers eight
cardiac cycles, and a periodic steady state is declared when the
pointwise $L_2$ difference between the last two cycles falls below
$10^{-3}$. 

Occlusions are imposed by removing the blocked branch and its
downstream subtree (or, for the ICA terminus, closing the distal ICA
with a total-reflection resistance); stenoses are represented
geometrically by a 1\,cm smooth converging--diverging throat of area
$A_1=(1-s_A)A_0$, which reproduces the viscous and inertial members of
equation~(\ref{eq:yt}) but not its quadratic term. All twelve scenarios of
Table~\ref{tab:scen}, plus a 55--75\,bpm heart-rate sweep, converged
without further intervention. The production comparison of
Section~\ref{sec:resval} was rerun end to end at 72\,bpm with the same solver, 
all twelve scenarios completing without
intervention.

\subsection{Three-dimensional stenosis case}
\label{sec:sv3d}

\begin{figure}[t]
\centering
\includegraphics[width=0.85\textwidth]{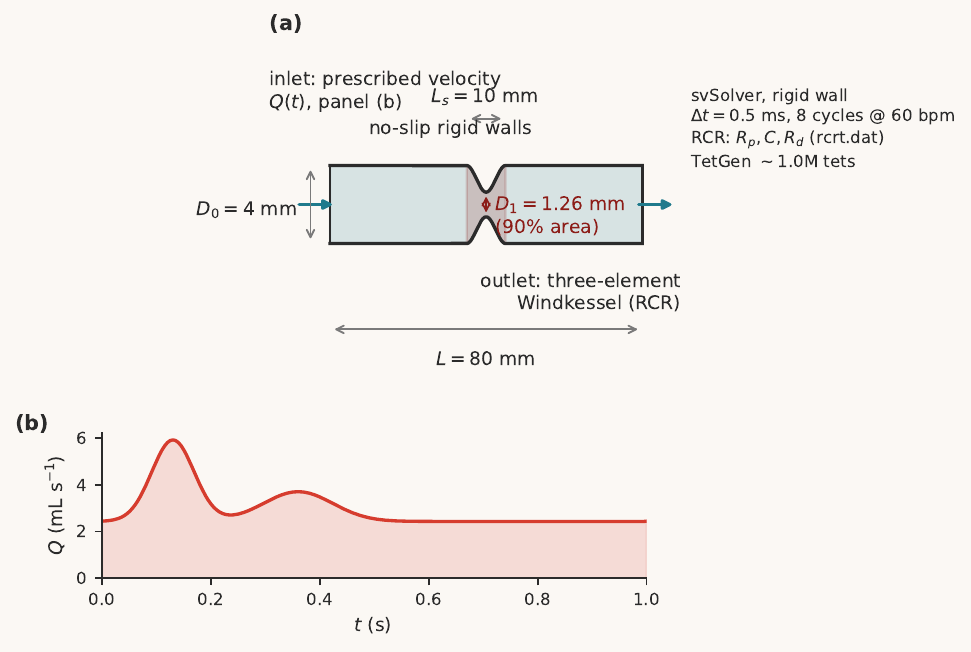}
\caption{{Computational domain and boundary conditions of
the three-dimensional stenosis validation case (90\,\% area reduction,
throat diameter $D_1=1.265$\,mm, 60\,bpm). (a) Idealised axisymmetric
ICA segment: a straight rigid tube of diameter $D_0=4$\,mm and length
$L=80$\,mm carrying a 10\,mm cosine throat (\ref{eq:throat}); inlet,
prescribed velocity interpolating the network model's carotid waveform
rescaled to the ICA mean flow of 2.97\,mL\,s$^{-1}$ (panel b); outlet,
three-element RCR Windkessel carrying the middle-cerebral bed parameters
of Section~\ref{sec:sv3d}; walls rigid and no-slip. (b) Prescribed
inflow rate over one cardiac cycle.}}
\label{fig:case3d}
\end{figure}
A second validation targets precisely the physics the linear theory
discards: the quadratic Young--Tsai loss and the spectral broadening of
the post-stenotic jet. The geometry (\textcolor{red}{Fig.~\ref{fig:case3d}}) is an
idealised axisymmetric ICA segment: a straight rigid tube of diameter
$D_0=4$\,mm and length $L=80$\,mm carrying a cosine-shaped stenosis of
length $L_s=10$\,mm centred mid-domain, with throat radius profile
\begin{equation}
r(z)=R_0-(R_0-R_1)\,\tfrac12\Bigl[1+\cos\frac{\pi(z-z_c)}{L_s/2}\Bigr],
\quad (|z-z_c|\le L_s/2),
\label{eq:throat}
\end{equation}
$R_0=D_0/2$, $z_c=L/2$ and $R_1/R_0=0.5$, i.e.\ a 75\,\% area reduction
matching the $s_A=75\%$ scenario of Table~\ref{tab:scen} and its
Young--Tsai coefficients. The diameter follows the carotid-bulb
experiments of Ojha et al.\ \citep{ojha1989}, which, with the channel
study of Mittal et al.\ \citep{mittal2003}, provide the profile,
reattachment and transition benchmarks for the computed flow.

{Three geometry choices deserve justification. The parent
diameter $D_0=4$\,mm is the calibre of the internal carotid segments of
the network model of Section~\ref{sec:model}---the verified parameter
set of \citep{alastruey2007} assigns $d=4.0$\,mm to both ICA
segments---and sits in the middle of the typical adult ICA lumen
diameter range. The domain length $L=80$\,mm ($=20\,D_0$) keeps the
inlet development length and the outlet recovery region well clear of
the throat's influence zone, so that the prescribed inflow is fully
developed before the converging section and the Windkessel outlet is not
exposed directly to the post-stenotic jet. The cosine throat profile
equation~(\ref{eq:throat}) of length $L_s=10$\,mm is the standard idealisation of
an atherosclerotic stenosis in lumped and distributed lesion models
\citep{young1973}.}

The incompressible Navier--Stokes equations are integrated with the
SimVascular finite-element solver svSolver \citep{updegrove2017}:
SUPG--PSPG stabilised Galerkin/least-squares formulation, second-order
generalised-$\alpha$ time integration with spectral radius
$\rho_\infty=0.5$, convective advection form, three nonlinear iterations
per step, residual control $10^{-2}$ and svLS tolerance $0.05$. The
tetrahedral mesh  uses a global edge length of
$D_0/20$ refined to $D_0/40$ across the throat, giving
$1.0\times10^{6}$ elements and $1.83\times10^{5}$ nodes. The time step is
$\Delta t=1$\,ms; 8{,}000 steps cover eight cardiac cycles at 60\,bpm
from a zero initial field. Blood is modelled as a Newtonian fluid of
density $1.06$\,g\,cm$^{-3}$ and viscosity $0.04$\,dyn\,s\,cm$^{-2}$; the
wall is rigid and no-slip. The inlet prescribes velocities interpolating
the network model's 201-point carotid waveform, rescaled to the ICA mean
flow of 2.97\,mL\,s$^{-1}$; the outlet is closed by a three-element RCR
Windkessel with $R_p=3.96\times10^{4}$\,dyn\,s\,cm$^{-5}$,
$C=1.16\times10^{-5}$\,cm$^{5}$\,dyn$^{-1}$ and
$R_d=2.01\times10^{4}$\,dyn\,s\,cm$^{-5}$ (CGS), the conversion of the
equivalent downstream load of the middle-cerebral territory in the
network model. 

{Specifically, the three values are the
CGS transcription of the middle-cerebral (M2) Windkessel load of the
one-dimensional network of Section~\ref{sec:model}, so that the
three-dimensional outflow condition is inherited from---not tuned
for---the same vascular bed used throughout this study. With
$1$\,Pa\,s\,m$^{-3}$ $=10^{-5}$\,dyn\,s\,cm$^{-5}$ and
$1$\,m$^{3}$\,Pa$^{-1}$ $=10^{5}$\,cm$^{5}$\,dyn$^{-1}$: the proximal
resistance matches the characteristic impedance of the terminal M2
segment, $R_p=R_1=\rho c_0/A_0=3.957\times10^{9}$\,Pa\,s\,m$^{-3}$
$\to3.96\times10^{4}$\,dyn\,s\,cm$^{-5}$ (proximal impedance matching);
the distal resistance is the remainder of the verified bed total
$R_{\mathrm{tot}}=5.97\times10^{9}$\,Pa\,s\,m$^{-3}$,
$R_d=R_{\mathrm{tot}}-R_1=2.01\times10^{9}$\,Pa\,s\,m$^{-3}$
$\to2.01\times10^{4}$\,dyn\,s\,cm$^{-5}$, which sets the mean outlet
pressure to the downstream microvascular level; and the compliance
$C=1.16\times10^{-10}$\,m$^{3}$\,Pa$^{-1}$
$\to1.16\times10^{-5}$\,cm$^{5}$\,dyn$^{-1}$ is the verified
middle-cerebral bed value.} 

The computation ran on eight MPI ranks of an Apple M1~Max
workstation (svSolver and its VTK/OpenMPI dependencies built from
source), advancing at $\simeq$3.5\,s per time step---about eight hours
of wall time for the eight cycles. Correctness was verified by (i) node-by-node inflow checks against the prescribed
velocity file, (ii) integrated-flow comparison (within two per cent in
the mean and nine per cent in the first three harmonics), and (iii) a
coarsened-mesh run with $0.58$ times the degrees of freedom under
identical conditions, which changed the peak throat velocity by only
0.5\,\%.

\begin{figure}[t]
\centering
\includegraphics[width=\textwidth]{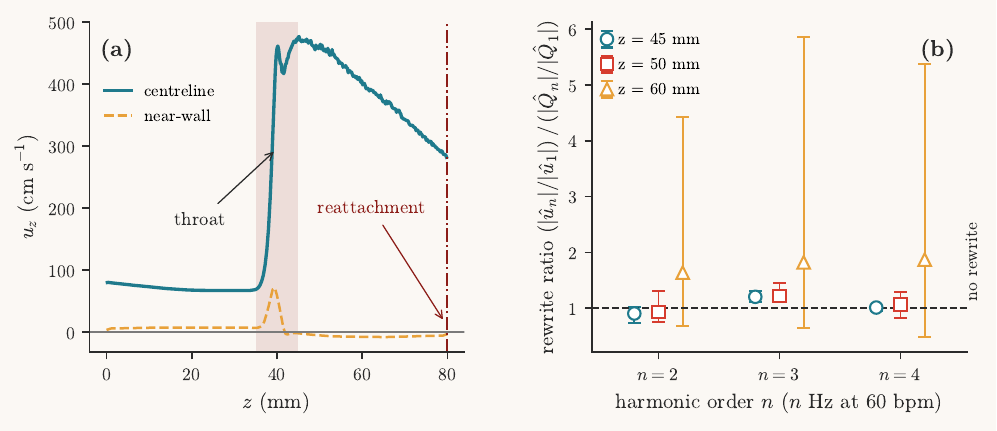}
\caption{{Three-dimensional stenosis results at 90\,\%
area reduction (throat diameter $D_1=1.265$\,mm, 60\,bpm). 
(a) Centreline (solid) and near-wall (dashed) axial velocity at peak
systole (cycle eight), with the stenosis bounds (shaded) and the
reattachment point of the post-stenotic separation bubble (dash-dotted).
(b) Harmonic rewrite ratio of the centreline velocity at the downstream
stations $z=45,50,60$\,mm relative to the inlet waveform for harmonics
$n=2,3,4$ (cycle means over cycles three to eight, whiskers the
cycle-to-cycle range; unity dashed).}}
\label{fig:3dresults}
\end{figure}
A second configuration tightens the stenosis to a 90\,\% area reduction,
$R_1=R_0\sqrt{0.1}$ (throat diameter $D_1=1.265$\,mm), matching the
$s_A=90\%$ scenario of Table~\ref{tab:scen} and shown in
{Fig.~\ref{fig:3dresults}}. The same meshing strategy gives $1{,}013{,}095$
tetrahedra and $184{,}579$ nodes, and the boundary conditions, fluid
properties and eight-cycle protocol are unchanged. The tighter throat
proved numerically stiffer: the nonlinear iteration diverged during the
first systolic upstroke at $\Delta t=1$\,ms and, with the step halved to
$\Delta t=0.5$\,ms (16{,}000 steps), again in the fifth-cycle upstroke,
each time exhausting the residual control. The completed run therefore
pairs $\Delta t=0.5$\,ms with stronger generalised-$\alpha$
high-frequency damping ($\rho_\infty=0.2$), restarted from the last
converged state, and advances the full eight cycles on the same eight
MPI ranks. We record the sequence because the stiffness is itself a
marker of the strongly chaotic jet quantified in
Section~\ref{sec:resval3d}.

   To quantify whether the lesion itself modifies the frequency content of                                
     the flow, we further define the {harmonic rewrite ratio} of harmonic $n$ at                               
     a downstream station $z$ as                                                                            
     \begin{equation}                                                                                       
       R_n(z) \;=\;                                                                                         
       \frac{|\hat{V}_n(z)|/|\hat{V}_1(z)|}                                                                 
            {|\hat{V}^{\mathrm{in}}_n|/|\hat{V}^{\mathrm{in}}_1|},                                          
       \label{eq:rewrite}                                                                                   
     \end{equation}                                                                                         
     where $\hat{V}_n$ is the complex Fourier amplitude of the centreline                                   
     velocity at the $n$-th cardiac harmonic and the superscript                                            
     $\mathrm{in}$ denotes the inlet waveform. A value of unity means that                                  
     the relative harmonic content survives the stenosis unchanged; values                                  
     above (below) unity indicate enrichment (damping) of that harmonic by                                  
     local jet physics. Because the linearised network model treats each                                    
     cardiac harmonic as an independent Womersley mode with no                                              
     inter-harmonic energy transfer, it predicts $R_n\approx1$ away from the                                
     lesion, so that equation~\eqref{eq:rewrite} constitutes a falsifiable check of                                  
     the linearisation against the fully nonlinear three-dimensional                                        
     computation.

\subsection{External benchmark}
\label{sec:benchproto}

The benchmark of \citep{alastruey2007} reports (i) healthy efferent flow
rates for the complete network and nine anatomical variants (absent ACoA,
absent/hypoplastic PCoA, absent A1 or P1, and combinations), and (ii) the
percentage flow change in the six efferent arteries following right ICA or
right VA occlusion in each variant, together with a communicating-artery
calibre sweep. We reproduced all three on the present network. Because the
two studies differ in boundary-condition philosophy---\citep{alastruey2007}
drive the network from a single cardiac (ascending-aorta) inflow, so an
occlusion redistributes a fixed cardiac output, whereas the present model
prescribes four independent inlet flows---each occlusion case was computed
in two variants: {frozen}, the native protocol with inlet flows held
at their healthy values, and {compensated} (comp), in which the
remaining inlets are rescaled to keep the total headward inflow constant
at its healthy value of 12\,mL\,s$^{-1}$, mimicking cardiac-output
redistribution. The two variants bracket the physiological truth; we
report both rather than choosing one. The calibre sweep varies the ACoA
(resp.\ PCoA) diameter over 0.2--3.0\,mm under the occlusion that engages
it.

\subsection{Virtual cohort and occlusion-localisation classifier}
\label{sec:cohort}

%\subsubsection{Cohort generation}
In this work, we generated three Monte Carlo simulated cohorts using the linear model. Cohort~A (standard geometry) consists of 800 jittered realisations per class for the twelve scenarios in Table~\ref{tab:scen} (healthy, eight occlusions, three M1 stenosis grades), totalling 9\,600 samples. Cohort~B (anatomical-variant test sets) consists of 100 realisations per class for each of six Circle of Willis variants: absent ACoA, absent left PCoA, absent right PCoA, absent bilateral PCoA, A1 hypoplasia, and P1 hypoplasia (hypoplasia modelled by halving the segment diameter). Cohort~C (geometry-augmented training set) provides 640 realisations per class, with the anatomical variant drawn uniformly at random per sample.

Inter-subject variability is modelled by perturbing the heart rate uniformly over 55--75~bpm; independently scaling each Windkessel parameter $R_2$ and $C$ by a uniform factor in $[0.8,\ 1.25]$; varying the viscosity by $\pm 10\,\%$; and perturbing the harmonic amplitudes and systolic phase of the inlet waveform. Every realisation is a fresh network solve, with per-sample costs in the millisecond range; consequently, generating the full cohort and running all subsequent experiments required 2{,}501\ \text{s} of single-core computation time.

\subsubsection{Feature sets}

Two feature families were compared, both derived from the velocity spectra at the midpoint of the CCA bilaterally. The \emph{physics-guided (PG) features} consist of the $\log_{10}$ harmonic-amplitude ratios against the nominal healthy model at the same heart rate (harmonics $n=1,\dots,10$, bilaterally; 20 dimensions), the ripple period and depth of the ratio spectrum (bilaterally; 4), and the velocity pulsatility index (PI; bilaterally; 2), yielding 26 dimensions in total. The healthy reference is obtained from the un-jittered standard-geometry model, serving as the nominal baseline that a clinician's digital twin would predict; inter-subject jitter thus manifests as scatter about unity in the ratio domain. To isolate the contribution of the ripple features, we also consider a ripple-free variant, PG$_{\mathrm{nr}}$, comprising only the ratios and PI (22 dimensions). 

The \emph{raw features}, by contrast, consist of the $\log_{10}$ harmonic amplitudes (20) plus the ripple period and depth of the measured spectrum (4), conveying the same information but without the healthy-model referencing. By construction, the PG features encode precisely the two signatures derived in Sections~\ref{sec:lesions}--\ref{sec:ripple}: the redistribution of harmonic ratios and the ripple of Eq.~(\ref{eq:ripple}).

\subsubsection{Classifiers and evaluation}

The primary classifier is linear discriminant analysis (LDA); a small
two-layer multilayer perceptron (64 hidden units, tanh, softmax,
full-batch Adam) serves as a nonlinear comparison. All evaluations use a
stratified 80/20 train/test split within cohort A (unless stated
otherwise). 

Four experiments were run. (1) \emph{Main accuracy}: 12-class
and a merged 10-class variant (single M1-stenosis class), with per-class
true-positive rates (TPR) and confusion matrices. (2) \emph{Noise
robustness}: white Gaussian noise of standard deviation
$\sigma$ times the waveform root-mean-square value is added to the
reconstructed time waveform (2{,}048 samples per beat) before harmonic
extraction, and multiplicative complex noise of the same relative level is
applied to the dense spectrum before the ripple fit; $\sigma$ ranges from
0 to 20\,\%. A noise-robust PI (mean of the top and bottom 1\,\% of
samples rather than max/min) is used throughout. Classifiers are evaluated
both trained clean and trained with noise augmentation (five noisy copies
per training sample at random $\sigma$). (3) \emph{Domain gap}:
classifiers trained on cohort A are tested on each variant of cohort B,
and the defence---training on the geometry-augmented cohort C---is
evaluated on the same targets. (4) \emph{Sample complexity}: accuracy as a
function of the number of training samples per class
(5--640).

%=====================================================================
\section{Results}
\label{sec:results}

\begin{figure}[t]
\centering
\includegraphics[width=\textwidth]{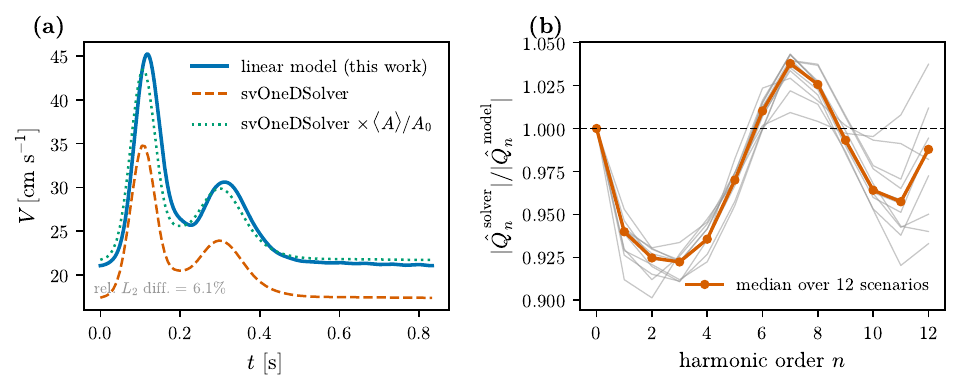}
\caption{Validation at the right common carotid midpoint station,
72\,bpm (all twelve scenarios of Table~\ref{tab:scen} rerun with the
arm64-native solver build). (a) Velocity waveform over one cardiac
period, healthy network: thin solid line, present linear frequency-domain
model; dashed line, nonlinear one-dimensional solver; dotted line, the
same solver trace rescaled by the mean wall-distension factor $\langle
A\rangle/A_0\approx1.2$ (relative $L_2$ difference after correction
6.1\,\%). (b) Ratio of harmonic flow magnitudes, nonlinear solver over
linear model, versus harmonic order: thin grey lines, all twelve
scenarios; thick line with circles, median. The ratios for $n=1$--$6$
span 0.901--1.024; the excursion above unity beyond the sixth harmonic is
the signature of nonlinear harmonic generation.}
\label{fig:validation}
\end{figure}
\subsection{Model validation by 1D solver}
\label{sec:resval}

%\subsubsection{Nonlinear one-dimensional solver}

Figure~\ref{fig:validation} summarises the comparison with the nonlinear
one-dimensional solver at the right CCA midpoint station across all twelve
scenarios at 72\,bpm. The direct-current (mean) flow divisions agree with
the linear model to three significant figures, and mean inlet pressures to
within three per cent, including the occlusion-induced pressure rise of
about $+15$\,mmHg for a middle-cerebral occlusion. Harmonic flow
magnitudes agree within a ratio of 0.90--1.04 for all harmonics $n\le12$
and all scenarios (0.901--1.024 for $n\le6$; medians $|Q_1|=0.940$,
$|Q_2|=0.925$), with the same systematic pattern observed at
60\,bpm---low harmonics a few per cent low, harmonics above the sixth a
few per cent high---that is the fingerprint of the nonlinear harmonic
generation deliberately discarded by the linear model, consistent with
the second-order estimate of Section~\ref{sec:lesions}. 

Velocity
magnitudes and pulsatility indices differ systematically by about twenty
and ten per cent respectively, a difference traced to the kinematic
conversion $V=Q/A_0$ used by the linear model: the compliant solver
operates at a mean distended area $\langle A\rangle/A_0\approx1.2$, and
rescaling by this factor collapses the waveforms onto each other
(Fig.~\ref{fig:validation}a, relative $L_2$ difference 6.1\,\% after the
correction). Because the discrepancy is a single side-independent factor,
it cancels in every harmonic-ratio and left--right-ratio signature used
below, and no retuning was attempted. This reduced-order premise has
precedent: a one-dimensional cerebral network coupled to lumped
intracranial compartments showed that extracranial velocity ratios
respond strongly and non-monotonically to graded vasospasm, with changes
above ten per cent within reach of transcranial Doppler \citep{ryu2017};
the present framework recovers the same non-monotone severity dependence
with a mechanistic reading---the severity dependence of the scattering
strength, i.e. equation~(\ref{eq:gammas}), competing against collateral re-routing of the
mean flow.

\subsection{Model validation by 3D solver}
\label{sec:resval3d}

The three-dimensional computation confirms both ends of the lumped lesion
representation ({Figs.~\ref{fig:case3d} and \ref{fig:3dresults}}). Exact periodicity is not attained: at
the peak throat Reynolds number $\mathrm{Re}_1\simeq10^{3}$ the
post-stenotic jet is weakly chaotic, same-phase velocity fields of
consecutive cycles differing by $10^{-2}$--$2\times10^{-1}$ in relative
$L_2$ norm throughout cycles four to eight (largest at the diastolic
frame, where velocities are smallest), while the cycle-mean quantities
quoted below vary by only a few per cent between cycles. At peak systole
the throat velocity reaches $194\pm7$\,cm\,s$^{-1}$, about sixty per cent of the quasi-steady
parabolic estimate $2Q/A_1\simeq318$\,cm\,s$^{-1}$---the five-millimetre
converging section is too short to establish a parabolic throat profile,
which is precisely why the lesion law retains the full three-term
Young--Tsai form, i.e. equation~(\ref{eq:yt}), rather than its quadratic member alone. 

A
separation bubble lines the diverging wall and reattaches
$2.6\pm0.6\,D_0$ downstream of the throat exit, of the order of the
several-diameter reattachment lengths measured experimentally at
comparable Reynolds number \citep{ojha1989}. The cycle-mean pressure drop
across the stenosis settles to 3.9\,mmHg, within ten per cent of the full
three-term Young--Tsai reconstruction driven by the measured flow rate
(3.6\,mmHg, of which the quadratic term contributes 2.2\,mmHg;
9.3\,mmHg at peak flow); instantaneous systolic drops fluctuate between 8
and 19\,mmHg with the weakly chaotic jet state. With the Windkessel load
active the cycle-mean outlet pressure is $143\pm11$\,mmHg, within ten per
cent of its theoretical periodic value $(R_p+R_d)\bar Q=131$\,mmHg, and
frame by frame the measured outlet pressure follows the Windkessel
differential equation driven by the measured outflow to within 9\,mmHg.
The post-stenotic harmonic content scatters about
that of the inlet waveform without systematic enrichment---rewrite ratios
of $n=2,3$ span 0.7--1.9 (cycle means 1.3--1.4) and $n=4$ is mildly
damped (0.6--1.6, cycle mean 0.9)---in line with the linear model's
prediction that the lesion rewrites the carotid-station harmonic ratios
by less than one per cent through local jet physics: the diagnostic
imprint is carried by network-scale scattering, not by spectral
broadening surviving downstream. Clinically, the computed peak-systolic-velocity ratio across
the lesion ($\approx2.5$ for this 75\,\% area, 50\,\% diameter reduction)
sits squarely on the empirical duplex threshold ``PSV ratio $>2$
$\Rightarrow$ $\ge50\%$ diameter stenosis'' compiled in \citep{nuffer2017},
anchoring the lumped representation to the velocity criteria used at the
bedside.

Tightening the lesion to 90\,\% area reduction ($D_1=1.265$\,mm;
{Fig.~\ref{fig:3dresults}}) intensifies every one of these signatures while
leaving the model hierarchy intact. The jet is now strongly chaotic:
same-phase velocity fields of consecutive cycles differ by
$0.25$--$0.52$ in relative $L_2$ norm, against $0.01$--$0.23$ at 75\,\%,
and cycle-mean quantities stabilise from the third cycle. At peak
systole the throat velocity reaches $483\pm9$\,cm\,s$^{-1}$, about sixty
per cent of the quasi-steady parabolic estimate
$2Q/A_1\simeq800$\,cm\,s$^{-1}$ ($Q/A_1\simeq400$\,cm\,s$^{-1}$)---the
same bluntness factor as at 75\,\%. The separation bubble reattaches
$8.6\pm0.4\,D_0$ downstream of the throat exit, more than three times
the 75\,\% value. The cycle-mean pressure drop rises to 26\,mmHg (cycle
range 25--27), ten to thirteen per cent below the full three-term
Young--Tsai reconstruction driven by the measured flow rate (29\,mmHg,
now overwhelmingly inertial: 77.7 of 79.6\,mmHg at peak flow);
instantaneous systolic drops fluctuate between 63 and 81\,mmHg about the
reconstructed 80\,mmHg with the jet state, and the measured and
reconstructed drops remain correlated frame by frame at
$r=0.97$--$1.00$. 

The harmonic rewrite ratios still scatter about unity
without systematic enrichment (cycle means 0.9--1.2 for $n=2,3,4$ at the
two probes nearest the throat, $n=5$ mildly damped at 0.7--0.8, and
isolated excursions up to $\sim$6 at the farthest station in the most
chaotic cycle), consistent with the linear model's prediction for the
$s_A=90\%$ scenario, whose carotid-station harmonic ratios stay within
two per cent of unity. The PSV ratio across the lesion rises to
$\approx6.2$, against $\approx2.5$ at 75\,\% and deep into the severe
side of the same duplex criterion \citep{nuffer2017}, while the measured
mean inflow ($2.92$\,mL\,s$^{-1}$) is essentially unchanged: even at
90\,\% area reduction the lesion remains near-transparent at the carotid
station, and its signature is carried by the local velocities it
multiplies.

Physically, rewrite ratios of order unity confirm that the diagnostic                                  
     imprint of an intracranial lesion is carried by network-scale                                          
     scattering---the occlusion acting as a reflector that re-organises the                                 
     standing-wave pattern of the circle of Willis---rather than by spectral                                
     broadening produced in the local jet that survives downstream. The                                     
     isolated excursions (up to $\sim6$) at the farthest station                                            
     $z=60$\,mm in the most chaotic cycle do not contradict this picture. In                                
     the post-reattachment recovery region the phase-locked harmonic                                        
     amplitudes are small after viscous decay and jet mixing, so the                                        
     normalised ratio \eqref{eq:rewrite} becomes ill-conditioned; broadband,                                
     non-phase-locked fluctuation energy of the weakly chaotic jet then leaks                               
     into the harmonic bins of the single-cycle Fourier transform and can                                   
     exceed the residual coherent amplitude. These excursions are sporadic                                  
     rather than cycle-coherent---they appear in the whiskers, not in the                                   
     cycle means, which remain near unity---they involve a minute absolute                                  
     energy compared with both the local fundamental and the carotid-station                                
     signature, and they are absent in the milder 75\,\% case. The linear                                   
     prediction concerns systematic, reproducible enrichment; stochastic                                    
     far-field fluctuations of this kind carry no usable diagnostic                                         
     information and do not reach the carotid measurement station.

\subsection{External benchmark results}
\label{sec:benchmark}

\begin{table}[t]
\centering
\caption{Healthy and collateral flow rates (mL\,s$^{-1}$) of the present
model against the benchmark of Alastruey et al.\ \citep{alastruey2007}
(their Table~2 and variant cases). R/LACA, R/LMCA, R/LPCA: right/left
anterior, middle and posterior cerebral arteries; ACoA, R/LPCoA:
communicating arteries (sign gives direction).}
\label{tab:bench}
\footnotesize
\begin{tabular}{lcccccc}
\toprule
Case & \multicolumn{2}{c}{RACA} & \multicolumn{2}{c}{RMCA} &
\multicolumn{2}{c}{RPCA} \\
\cmidrule(lr){2-3}\cmidrule(lr){4-5}\cmidrule(lr){6-7}
 & present & ref.\ & present & ref.\ & present & ref.\ \\
\midrule
complete network & 1.257 & 1.16 & 1.749 & 1.73 & 0.967 & 0.90 \\
absent right A1, ACoA flow & \multicolumn{6}{l}{1.19 vs 1.07 ($+11\%$)} \\
absent right P1, PCoA flow & \multicolumn{6}{l}{$-0.91$ vs $-0.79$ ($+16\%$)} \\
\bottomrule
\end{tabular}
\end{table}

\begin{figure}[h!]
\centering
\includegraphics[height=0.8\textheight]{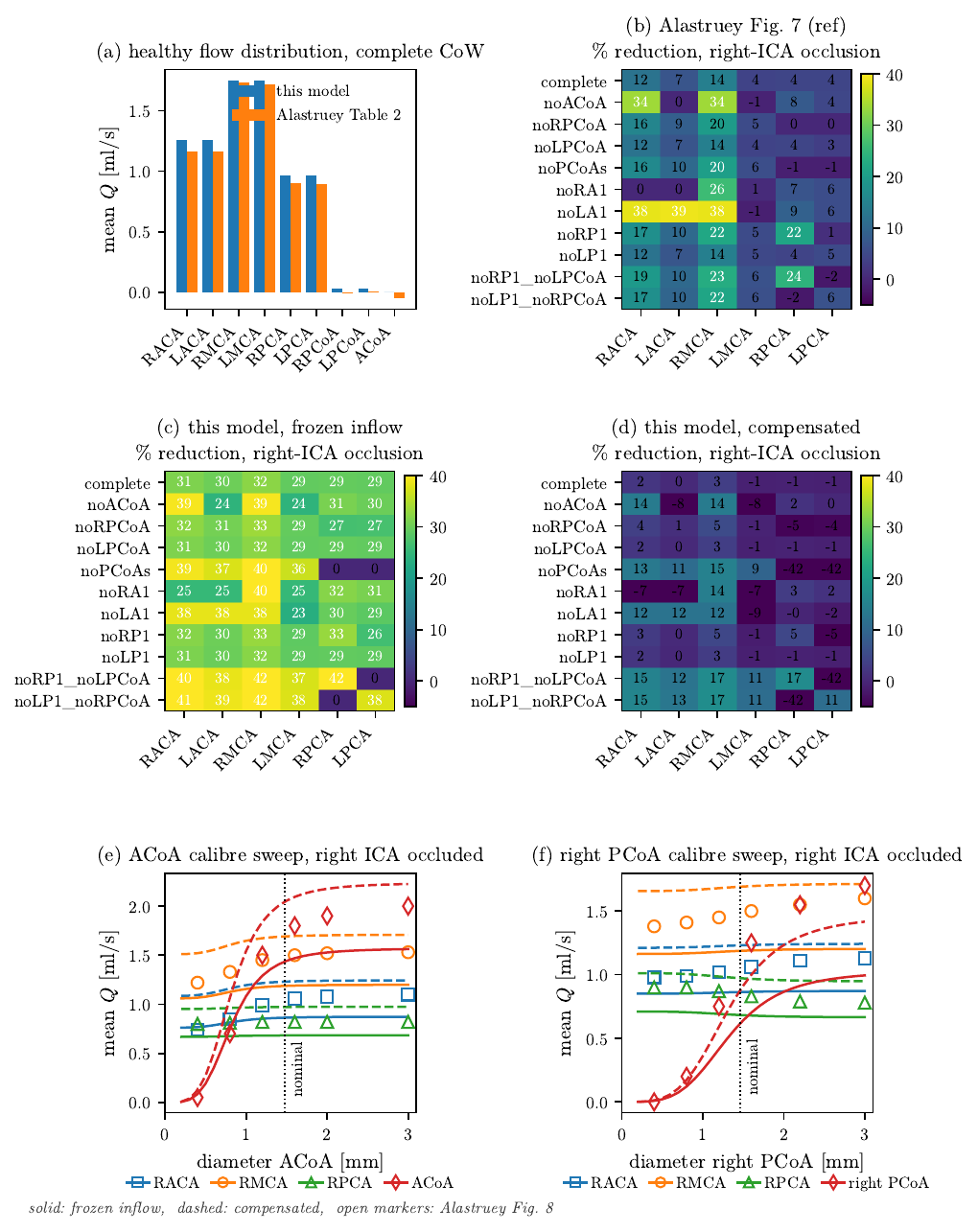}
\caption{External benchmark against Alastruey et al.\ \citep{alastruey2007}.
(a) Healthy efferent flow rates, present model versus reference. {(b--d)}
Percentage flow reduction in the six efferent arteries after right ICA
{occlusion} for each anatomical variant: {reference (b), frozen-inflow (c)
and compensated-inflow (d)} variants of the present model. {(e, f)}
Collateral-channel flow versus communicating-artery calibre under the
occlusion that engages it: ACoA {(e)} and PCoA {(f)}; symbols mark the
reference values at the calibres they report.}
\label{fig:benchmark}
\end{figure}
Figure~\ref{fig:benchmark} and Table~\ref{tab:bench} summarise the
external benchmark. For the healthy complete network, the efferent flow
rates agree with \citep{alastruey2007} to within 1--9\,\%: RACA 1.257 vs
1.16 ($+8.4\%$), RMCA 1.749 vs 1.73 ($+1.1\%$), RPCA 0.967 vs 0.90
($+7.4\%$)\,mL\,s$^{-1}$. The collateral channels that the anatomical
variants engage are likewise reproduced in pathway, direction and order of
magnitude: with the right A1 absent, the ACoA carries 1.19 vs
1.07\,mL\,s$^{-1}$ ($+11\%$); with the right P1 absent, the ipsilateral
PCoA reverses to $-0.91$ vs $-0.79$\,mL\,s$^{-1}$ ($+16\%$).

For the occlusion cases, the two boundary-condition philosophies
(Section~\ref{sec:benchproto}) matter more than any model error, and the
benchmark values fall systematically between the frozen and compensated
variants (Fig.~\ref{fig:benchmark}{b--d}). For a right ICA occlusion on
the complete network, the RMCA flow reduction is 32.1\,\% frozen, 3.3\,\%
compensated, against the reference 14.5\,\%; with the ACoA absent the
corresponding numbers are 39.4\,\%, 13.7\,\% and 33.5\,\%. For a right VA
occlusion the compensated variant agrees well with the reference
(RPCA/LPCA reductions of 1.9\,\% against 3.3\,\%). 

In summary, the qualitative
hierarchy of \citep{alastruey2007} is reproduced: the ACoA is the more
critical collateral pathway---its absence roughly doubles the ipsilateral
flow deficit under ICA occlusion in both variants---and the combination of
an absent A1 with a contralateral ICA occlusion is among the worst
configurations: the present model predicts reductions of 38.4\,\% in RACA,
LACA and RMCA, numerically coinciding with the reference values 38.4,
38.9 and 38.5\,\%. 

We should admit that one discrepancy remains: the combination of an
absent right P1 with an absent left PCoA is equally poor in the present
model (RACA $-40.3\,\%$, RMCA $-41.5\,\%$ frozen) but not in the
reference. We attribute this to the linear model's lack of the quadratic
pressure drop across the communicating segments, which penalises
high-flow collateral routes in the nonlinear model, together with the
frozen vertebral inflow (Section~\ref{sec:discussion}). The calibre sweep
(Fig.~\ref{fig:benchmark}{e, f}) reproduces the threshold behaviour of
\citep{alastruey2007}: the ACoA begins to carry collateral flow at
0.2--0.4\,mm diameter and the PCoA at about 0.4\,mm, and at the nominal
calibres (ACoA 1.48\,mm, PCoA 1.46\,mm) the collateral flows bracket the
reference---ACoA 1.44 frozen / 2.06 compensated vs $\approx$1.75, PCoA
0.64 / 0.91 vs $\approx$1.15\,mL\,s$^{-1}$.

\subsection{Spectral signatures}
\label{sec:spectra}

%\subsubsection{Harmonic signatures and the 2--12\,Hz band}

\begin{figure}[t]
\centering
\includegraphics[width=0.9\textwidth]{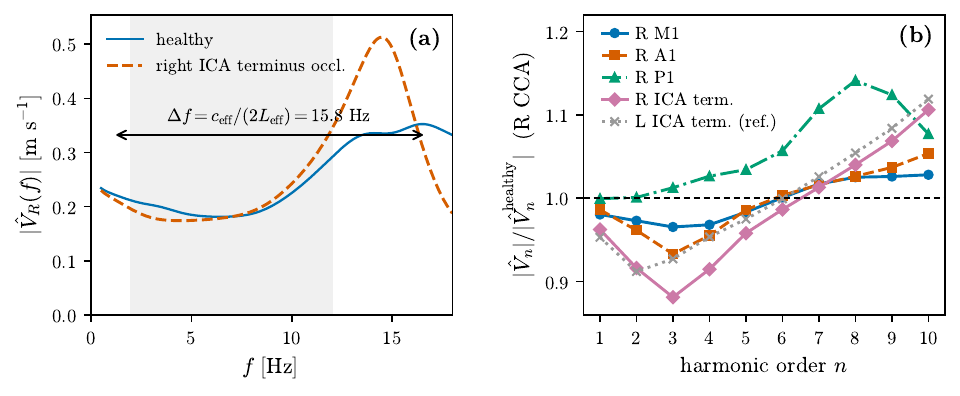}
\caption{Spectral signatures at the right common carotid station,
65\,bpm. (a) Dense transfer spectrum $|\hat V(f)|$ per unit inlet
spectrum: thin solid line, healthy; dashed line, occlusion of the right
ICA terminus. The grey band marks the empirical 2--12\,Hz discriminative
band; the double arrow marks the ripple spacing $\Delta
f=c_{\mathrm{eff}}/(2L_{\mathrm{eff}})=15.8$\,Hz predicted by
(\ref{eq:deltaf}) for this occlusion, which exceeds the entire band.
(b) Occlusion-to-healthy harmonic amplitude ratios versus harmonic
order, right side: right M1 (circles), right A1 (squares), right P1
(triangles) and right ICA-terminus (diamonds) occlusions; the faint
dotted line with crosses is the mirror left ICA-terminus occlusion at
the left station for reference. The minimum of the carotid-terminus
curve sits at $n=3$, one harmonic later than for the mirror left-sided
occlusion ($n=2$), reflecting the shorter right-sided reflection path.}
\label{fig:spectra}
\end{figure}

The healthy baseline reproduces physiological values at the carotid
station: mean velocity 0.255\,m\,s$^{-1}$, pulsatility index 0.92, mean
pressure 90\,mmHg, and a total cerebral perfusion of
476\,mL\,min$^{-1}$ against a prescribed inflow of 720\,mL\,min$^{-1}$.
Fifty-two per cent of the inlet-flow spectral energy sits in the
fundamental harmonic and 47.6\,\% in harmonics two to twelve.
Figure~\ref{fig:spectra} shows the dense transfer spectrum at the right
carotid station and the occlusion-to-healthy harmonic amplitude ratios
for the right-sided lesions (right M1/A1/P1 and ICA terminus) at 65\,bpm.
The signatures are structured and side-resolving: a right
carotid-terminus occlusion attenuates the ipsilateral low harmonics to
0.963, 0.916, 0.881, 0.915 and 0.958 of their healthy values for
$n=1,\dots,5$---the minimum sitting at $n=3$, one harmonic later than for
the mirror left-sided occlusion (minimum at $n=2$), because the shorter
right-sided path (right CCA 0.177\,m versus 0.208\,m left) pushes the
reflection dip to a slightly higher frequency---before the ratios recover
through unity near $n\approx6$ and amplify to 1.11 by $n=10$. 

A right P1
occlusion instead lifts the ratios gently \emph{above} unity from $n=3$
(up to 1.04 within $n\le5$), a sign-opposite, weaker signature
consistent with the posterior territory's small share of the carotid
flow. The collateral pathway activated by an M1 occlusion---posterior
communicating flow reversing from $-0.03$ to $+0.30$\,mL\,s$^{-1}$ and
anterior communicating flow rising from 0 to 0.83\,mL\,s$^{-1}$---is the
same communicating-artery recruitment that patient-specific computations
resolve in anatomical detail during vasospasm \citep{straccia2023}.

\begin{figure}[t]
\centering
\includegraphics[width=0.95\textwidth]{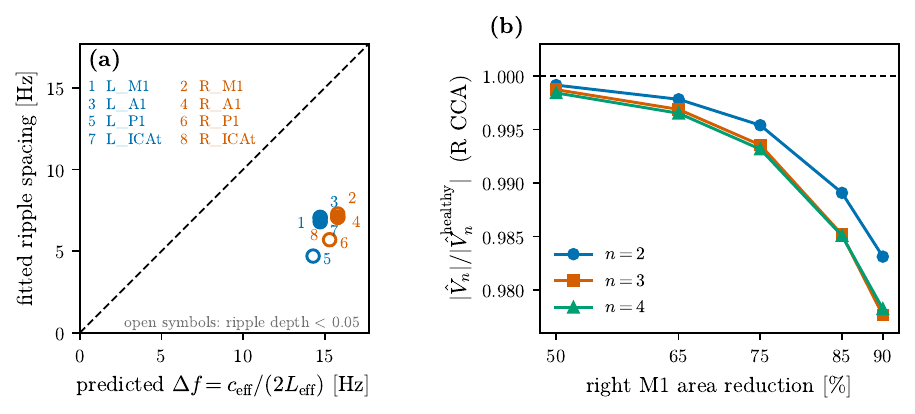}
\caption{Closed-form signatures against their in-band realisations,
65\,bpm. (a) Predicted ripple spacing $\Delta
f=c_{\mathrm{eff}}/(2L_{\mathrm{eff}})$ of (\ref{eq:deltaf}) versus the
spacing fitted within the 2--12\,Hz band at the ipsilateral CCA station
for the eight occlusion scenarios: 1/2 left/right M1, 3/4 left/right A1, 5/6
left/right P1, 7/8 left/right ICA terminus: all points lie far off the diagonal
(dashed), because the predicted spacings (14.3--15.8\,Hz, stratified by
side) fall outside the analysis band and the in-band fit recovers the
healthy standing-wave texture (4.7--7.3\,Hz). Open symbols mark ripple
depths below 0.05 (the two P1 occlusions and the right M1 occlusion). (b) Ipsilateral
occlusion-to-healthy harmonic ratios at $n=2,3,4$ versus M1 stenosis
grade (50--90\,\% area reduction): monotone, accelerating decay with
severity.}
\label{fig:signatures}
\end{figure}

For the eight occlusion scenarios the predicted ripple spacing
(\ref{eq:deltaf}) falls between 14.3 and 15.8\,Hz, stratified by side
(14.3--14.7\,Hz left, 15.3--15.8\,Hz right, the offset again a
consequence of the shorter right CCA; 15.8\,Hz for the right
carotid-terminus occlusion), entirely above the empirical 2--12\,Hz
discriminative band reported by data-driven studies
\citep{sen2024, argilaga2026} (the double arrow in
Fig.~\ref{fig:spectra}a spans more than the band itself).
Figure~\ref{fig:signatures}a makes the mismatch explicit: plotted
against the spacing actually fitted within the 2--12\,Hz band
(4.7--7.3\,Hz), every scenario falls far off the diagonal, because the
theoretical spacing lies outside the analysis band and the in-band fit
locks onto the standing-wave texture of the healthy network
($\simeq$7\,Hz), which exists in the healthy network as well and does
not move under occlusion. 

What carries occlusion information in the band
is the ripple \emph{depth}---spanning 0.02--0.11 across scenarios,
deepest for the carotid-terminus occlusions (0.075--0.112) and below
0.05 for the P1 occlusions and the right M1 occlusion (open symbols in
Fig.~\ref{fig:signatures}a), where the collateral supply masks the
echo---and the harmonic-ratio redistribution of Fig.~\ref{fig:spectra}b. The band itself is selected
by three physical factors: from below, the concentration of the cardiac
excitation spectrum in the fundamental and first few harmonics; from
above, the Womersley attenuation growing as $\sqrt{\omega}$ jointly with
the falling excitation; and the frequency dependence of the scattering
strength (\ref{eq:gammas}), largest at low frequencies. Because both the
excitation and the harmonic lattice scale with the heart rate, the band
is fundamentally a statement about \emph{harmonic orders}, not absolute
frequencies---2--12\,Hz is simply $n=2$--$12$ at 60\,bpm---and our
55--75\,bpm heart-rate sweep confirms that the signature envelopes are
stable when plotted against harmonic order and drift when plotted against
absolute frequency. We therefore recommend that data-driven studies
express discriminative bands in heart-rate-invariant harmonic orders
$n=f/f_{\mathrm{heart}}$ rather than in hertz.

Severity is encoded just as systematically. Figure~\ref{fig:signatures}b
scans the M1 stenosis grade $s_A=50/65/75/85/90\%$: the ipsilateral
harmonic ratios decrease monotonically with grade, from 0.999 to 0.983
at $n=2$ and to 0.978 at $n=3$ and $n=4$, the descent accelerating at
the tightest grade. The stenosis severity is thus written into the
carotid spectrum as a monotone redistribution of harmonic amplitude
ratios---small in absolute terms, but ordered---which is what allows the
classifier of Section~\ref{sec:classifier} to treat the stenosis grade
itself as a learnable class rather than a detection threshold.

\subsection{Synthetic spectral Doppler}

\begin{figure}[t]
\centering
\includegraphics[width=0.9\textwidth]{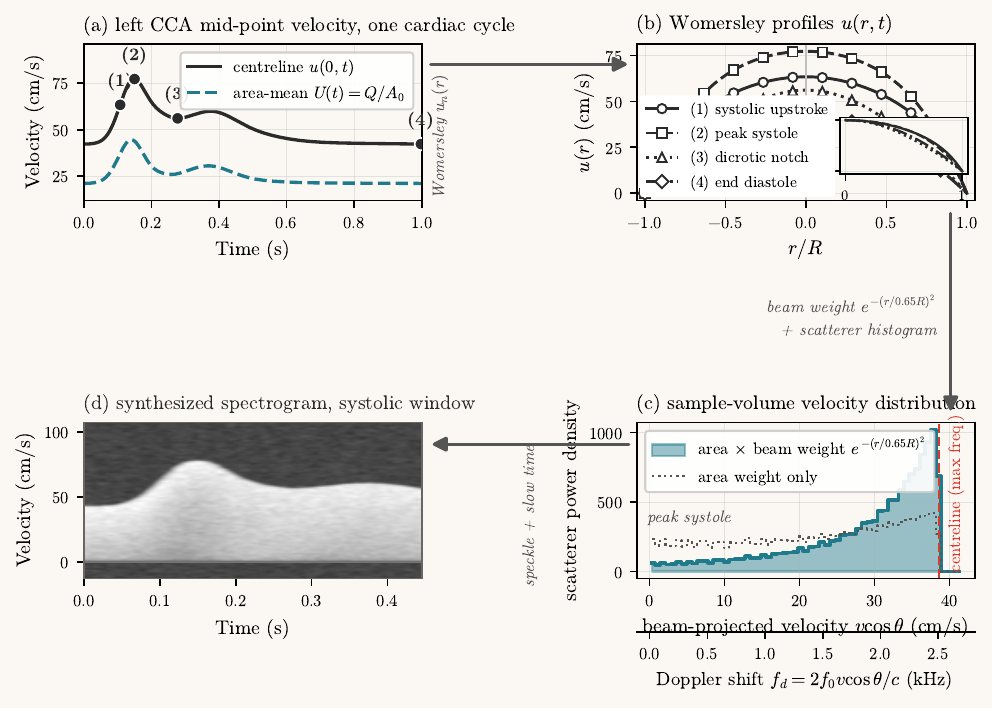}
\caption{The synthetic spectral-Doppler pipeline, step by step (healthy network, left CCA
midpoint). (a) CCA velocity waveform over one cardiac cycle. (b)
Womersley cross-sectional velocity profiles $u(r,t)$ at four phases of
the cycle: blunt and piston-like in systole, more nearly parabolic in
diastole. (c) Beam-weighted sample-volume velocity distribution at peak
systole: annulus-area and focused-beam weighting of the projected
profile. (d) Magnified view of the synthesised spectrogram around the
systolic window, showing the sub-systolic spectral window opened by the
blunt profile and beam weighting.}
\label{fig:dopplerpipeline}
\end{figure}
\subsubsection{The pipeline}
\label{sec:doppler}

To connect the network model to the clinical observable, the computed
harmonic velocity field at the CCA midpoint is converted into a
scanner-like spectral-Doppler image by a physics-based pipeline, worked
through step by step in Fig.~\ref{fig:dopplerpipeline}. For each
cardiac harmonic, Womersley's solution (\ref{eq:FJ}) supplies the full
cross-sectional velocity profile $u(r,t)$, which is reconstructed in time
by Fourier synthesis (Fig.~\ref{fig:dopplerpipeline}a, b: the profile is
blunt and piston-like in systole, when the high Womersley number
concentrates the oscillatory boundary layer at the wall, and more nearly
parabolic in diastole). The measured power at Doppler shift $f_d$ is the
scatterer histogram of the beam-projected velocity $u(r,t)\cos\theta$,
weighted by annulus area $2\pi r\,\dd r$ and by a focused-beam intensity
profile $\exp(-(r/0.65R)^2)$ centred on the lumen
(Fig.~\ref{fig:dopplerpipeline}c)---the blunt systolic
Womersley profile combined with the beam weighting is what opens the
sub-systolic spectral window in systole and the spectral broadening in
diastole, rather than an imposed artefact. 

Shifts are mapped to velocities
by the Doppler equation $f_d=2f_0 v\cos\theta/c_s$ with transmit frequency
$f_0=5$\,MHz, insonation angle $\theta=60^\circ$, sound speed
$c_s=1540$\,m\,s$^{-1}$ and pulse-repetition frequency 7\,kHz (Nyquist
3.5\,kHz). The spectrogram is then degraded realistically: fully developed
speckle (exponential power fluctuations), a noise floor, a mild 60\,Hz
wall filter suppressing vessel-wall clutter, scanner-like persistence
smoothing, and logarithmic (dB) compression to a clinical greyscale map.
The spectral envelope is extracted from the processed spectrogram by a
$-20$\,dB threshold, from which peak-systolic velocity (PSV),
end-diastolic velocity (EDV) and the resistivity index
$\mathrm{RI}=(\mathrm{PSV}-\mathrm{EDV})/\mathrm{PSV}$ are read exactly as
at the bedside. The same pipeline evaluates the stenotic M1 throat jet by
advecting the network flow rate through the throat area. The entire
pipeline costs seconds per scenario---orders of magnitude below
full-physics ultrasound simulation---and its harmonic-content output
agrees with the network solve to thirteen significant digits, so it is a
faithful rendering layer, not an additional approximation.

{More specifically, the synthesis of
Fig.~\ref{fig:dopplerpipeline}d proceeds in six steps: (i) the network
solve yields the complex area-mean velocity harmonics $V_n$
($n=0,\dots,25$) at the CCA midpoint; (ii) each harmonic is given its
Womersley cross-sectional profile
$u_n(r)=U_n\,[1-J_0(s_n r/R)/J_0(s_n)]/[1-F_J(s_n)]$ (parabola at
$n=0$) and the superposition is Fourier-synthesised into $u(r,t)$ over
one cardiac cycle; (iii) at each slow-time instant the scatterer
histogram of the beam-projected velocity $u(r,t)\cos\theta$ is weighted
by annulus area $2\pi r\,\dd r$ and by the focused-beam intensity
$\exp(-(r/0.65R)^2)$, giving the sample-volume velocity distribution of
panel (c); (iv) velocities are mapped to Doppler shifts by
$f_d=2f_0 v\cos\theta/c_s$ with $f_0=5$\,MHz and $\theta=60^\circ$;
(v) the power map is degraded by fully developed exponential speckle, a
$-40$\,dB noise floor, a mild 60\,Hz wall filter and scanner-like
persistence smoothing, and compressed logarithmically over a 45\,dB
dynamic range to a clinical greyscale map; (vi) the processed columns
are concatenated along slow time at the 7\,kHz pulse-repetition
frequency over three cardiac cycles with a $\sim$1\,\% cycle-to-cycle
gain jitter, producing the spectrogram of panel (d).}

\subsubsection{Spectral Doppler results}
\begin{figure}[!t]
\centering
\includegraphics[height=0.7\textheight]{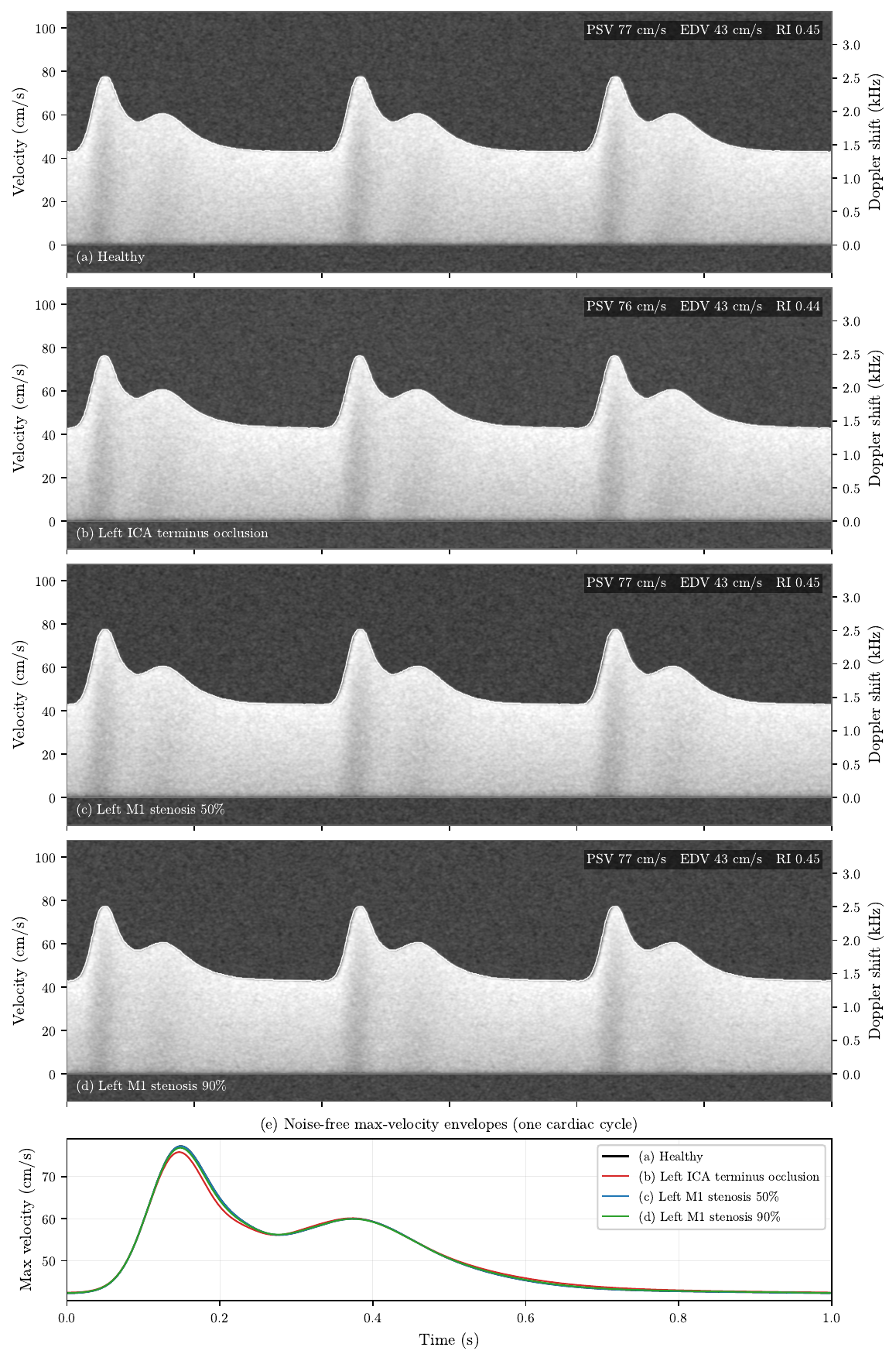}
\caption{Synthetic spectral-Doppler spectrograms at the left common
carotid station ($f_0=5$\,MHz, $\theta=60^\circ$, PRF 7\,kHz, 60\,bpm{;
three cardiac cycles per panel}).
(a) Healthy; (b) left ICA-terminus occlusion; (c, d) left M1 stenosis of
50\,\% and 90\,\% area reduction{; the scenario labels are
placed in the lower-left corner of each panel, and the PSV/EDV/RI values
read off the extracted envelope are annotated in the upper right}. The white trace is the envelope
extracted by a $-20$\,dB threshold. (e) {Noise-free
maximum-velocity (centreline) envelopes} of the four
scenarios superimposed {over one cardiac cycle}. Note the sub-systolic spectral window and the
diastolic spectral broadening, both produced by the Womersley profile and
focused-beam weighting.}
\label{fig:doppler}
\end{figure}
As an example, Fig.~\ref{fig:doppler} shows the synthetic spectrograms at the left CCA
station for the healthy network and three lesions, with the extracted
envelopes overlaid and superimposed. All four panels are
produced by the single pipeline of Section~\ref{sec:doppler}
(Fig.~\ref{fig:dopplerpipeline}) without per-scenario tuning: the solved
harmonics at the left CCA midpoint drive the Womersley-profile,
beam-weighted synthesis, each panel renders three cardiac cycles, and
the maximum-frequency envelope is extracted from the processed
spectrogram by a $-20$\,dB threshold relative to the column maximum
(with a $\sim$15\,ms median smoothing), from which PSV, EDV and RI are
computed exactly as at the bedside. With the per-panel titles removed in
the revised figure, the scenario labels sit in the lower-left corner of
each panel and the extracted PSV/EDV/RI values are annotated in the
upper right; panel (e) superimposes the noise-free maximum-velocity
envelopes of the four scenarios over one cardiac cycle.

The healthy trace is physiological:
PSV 77.4\,cm\,s$^{-1}$, EDV 42.7\,cm\,s$^{-1}$, RI 0.448, with a clean
sub-systolic spectral window in systole and diastolic spectral
broadening---both generated by the Womersley profile and beam weighting
rather than added by hand. A left ICA-terminus occlusion changes the CCA
envelope only weakly (PSV 76.3\,cm\,s$^{-1}$, $-1.4\%$): the
discriminative information is carried not by the envelope extrema but by
the pulsatility and the harmonic content---the PI falls from 0.921 to
0.883 and the ipsilateral second harmonic drops by 8.4\,\%, in exact
(thirteen-significant-digit) agreement with the network solve of
Section~\ref{sec:spectra}. M1 stenoses are nearly invisible at the carotid
station in terms of PSV ($-0.5\%$ even at $s_A=90\%$), consistent with the
linear theory's prediction that the local lesion rewrites the proximal
harmonic ratios by less than one per cent; the clinically meaningful PSV
elevation appears at the stenotic jet itself, where the computed M1 throat
PSV reaches 165.1\,cm\,s$^{-1}$ at $s_A=90\%$---a factor 4.58 over the
healthy M1 segment value of 36.0\,cm\,s$^{-1}$, and 70.3\,cm\,s$^{-1}$
(ratio 1.95) at $s_A=50\%$---in qualitative agreement with the bedside
PSV-ratio$>2$ grading criterion \citep{nuffer2017}.

\subsection{Occlusion-localisation classifier}
\label{sec:classifier}

\subsubsection{Why the physics-guided features work}
\label{sec:featdemo}

\begin{figure}[!tb]
\centering
\includegraphics[width=\textwidth]{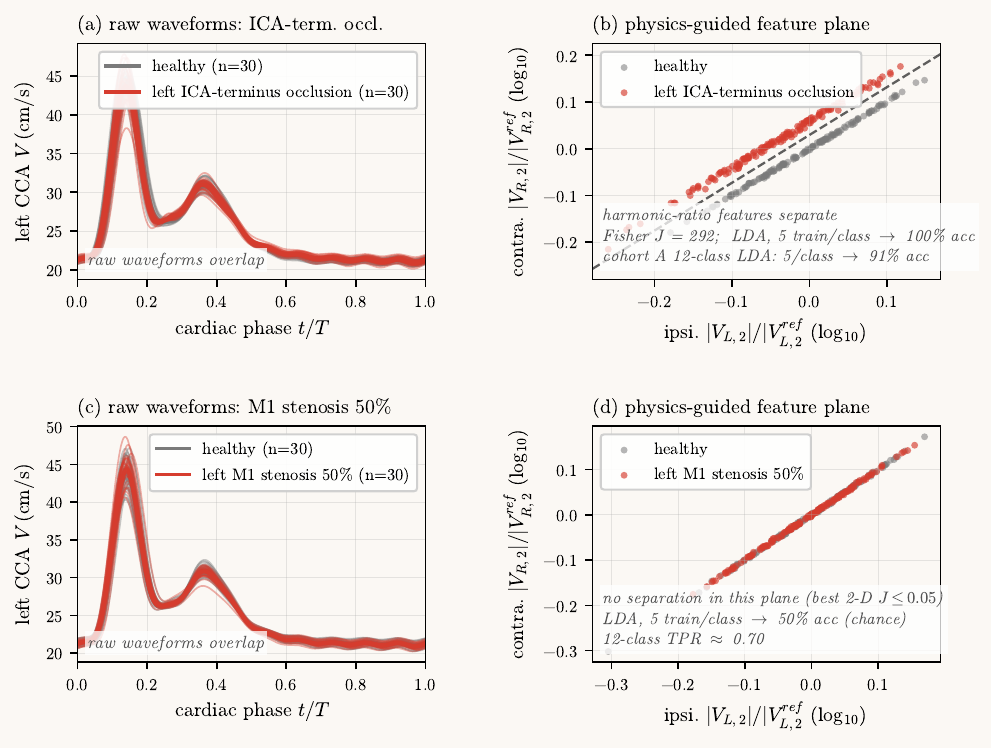}
\caption{Physics-guided features versus raw waveforms on two extreme
class pairs (cohort A, left CCA). (a, c) Overlaid raw velocity waveforms
(individual realisations and class means): indistinguishable by eye for
both pairs. (b) Second-harmonic amplitude-ratio feature plane
(ipsilateral versus contralateral, relative to the healthy reference
model) for healthy versus left ICA-terminus occlusion: Fisher
$J=291.7$; a two-class LDA trained on five samples per class reaches
100\,\% held-out accuracy (200 random training draws; dashed line, one
example boundary). (d) The same plane for healthy versus 50\,\% M1
stenosis: no separation ($J=0.003$, chance level; no 2-D plane of the
26-feature set exceeds $J=0.05$), the pair being separable only
partially in higher-dimensional feature combinations.}
\label{fig:featuredemo}
\end{figure}

As a demonstration, 
Fig.~\ref{fig:featuredemo} shows, on two extreme class pairs,
why the physics-guided representation is so sample-efficient. The raw
left-CCA waveforms of healthy subjects and of left ICA-terminus
occlusions are visually indistinguishable when overlaid
(Fig.~\ref{fig:featuredemo}a), yet in the physics-guided feature
plane---the ipsilateral versus contralateral second-harmonic amplitude
ratios against the healthy reference model---the two classes are cleanly
separated (Fig.~\ref{fig:featuredemo}b): the Mahalanobis-squared Fisher
separation is $J=291.7$, and a two-class LDA trained on only five
samples per class attains 100\,\% accuracy on 200 held-out samples per
class across 200 random training draws. The opposite extreme is healthy
versus the mildest M1 stenosis ($s_A=50\%$): the raw waveforms again
overlap (Fig.~\ref{fig:featuredemo}c), but here the same feature plane
shows no separation either (Fig.~\ref{fig:featuredemo}d, $J=0.003$,
chance level), and no two-dimensional projection of the full 26-feature
set does better (best 2-D plane $J\le0.05$). 

We should admit that this class
pair is only partially separable in higher-dimensional combinations of
the features---which is exactly where the twelve-class classifier's
residual errors concentrate (TPR $\approx0.70$ for M1 50\,\%,
Section~\ref{sec:classmain}). The demonstration thus brackets the claim
of this section: where the physics writes a signature, a handful of
samples suffice; where it does not, no feature engineering within the
carotid spectrum will manufacture one.

\subsubsection{Main results and sample complexity}
\label{sec:classmain}

\begin{figure}[!t]
\centering
\includegraphics[width=\textwidth]{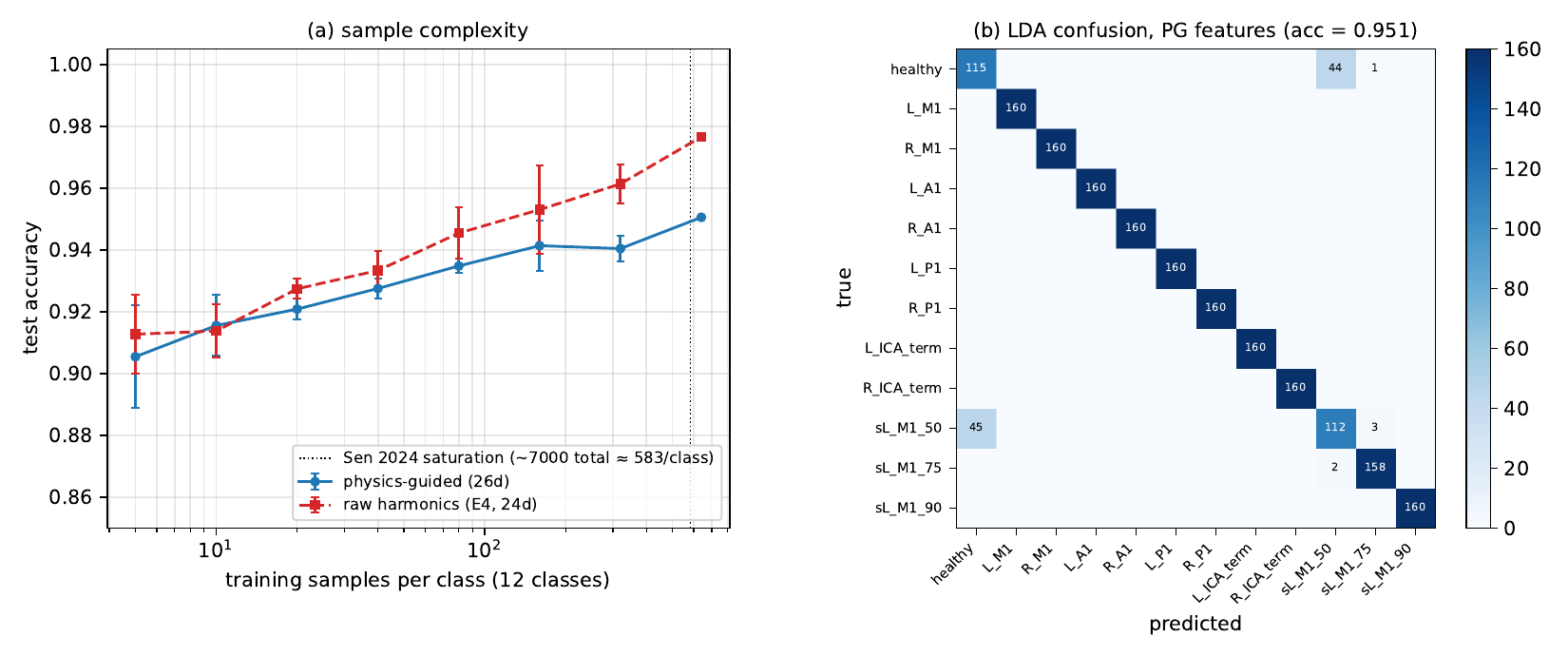}
\caption{Physics-guided occlusion localisation, cohort A. (a) Accuracy
versus the number of training samples per class for PG and raw features:
five samples per class already give 0.91, and saturation ($\approx$0.95,
PG) is reached from about 640 samples per class. (b) Twelve-class
confusion matrix of the LDA classifier on PG features: all eight occlusion
classes are detected with TPR 1.00; the residual confusion is between
healthy and the 50\,\% M1 stenosis.}
\label{fig:confusion}
\end{figure}
Table~\ref{tab:classifier} and Fig.~\ref{fig:confusion} summarise the main
classification experiment. With physics-guided features, LDA localises the
twelve classes at 95.1\,\% accuracy; raw amplitude features reach 97.7\,\%
and the small MLP on PG features 98.8\,\%. The per-class TPR of the PG
classifier is 1.00 for all eight occlusion classes; its residual errors
concentrate, honestly, where the physics says they should: healthy
(0.72) and the mildest stenosis grade (M1 50\,\%, 0.70) are mutually
confused, because a 50\,\% M1 stenosis rewrites the carotid spectrum at
the sub-per-cent level (Section~\ref{sec:spectra}), while the 75\,\% and
90\,\% grades are detected at 0.99 and 1.00. 

The decisive advantage of the
physics-guided representation is its much reduced sample complexity
(Fig.~\ref{fig:confusion}a): five training samples per class already give
90.5\,\% accuracy, and performance saturates at 95\,\% from about 640
samples per class---more than an order of magnitude below the
$\sim$7{,}000-sample saturation of waveform-driven convoluted neural networks (CNNs) on the same task
family \citep{sen2024}. Figure~\ref{fig:featuredemo}b is the microscopic
explanation of that curve: in the physics-guided plane the well-signed
class pairs are separated by Fisher distances of order $10^{2}$, so five
samples per class already pin down the class means and pooled covariance
of a two-class LDA to 100\,\% held-out accuracy; the residual
twelve-class error is carried by the physically near-degenerate pair of
Fig.~\ref{fig:featuredemo}d, which no training-set size can repair. The
entire cohort generation and all experiments
consumed 2{,}501\,s of single-core compute, because the linear forward
model costs milliseconds per sample; the nonlinear one-dimensional solver
underlying \citep{sen2024} costs of order $10^{3}$\,s per waveform, which
is precisely what forces the ML-surrogate strategy there.

\begin{table}[t]
\centering
\caption{Twelve-class localisation accuracy (stratified 80/20 split,
cohort A, 9{,}600 samples). PG: physics-guided features (harmonic-amplitude
ratios versus the nominal healthy model $+$ ratio-spectrum ripple $+$ PI);
PG$_{\mathrm{nr}}$: PG without ripple features; raw: unreferenced harmonic
amplitudes $+$ ripple.}
\label{tab:classifier}
\begin{tabular}{lcc}
\toprule
Classifier & Features & Accuracy \\
\midrule
LDA & PG (26\,d) & 0.951 \\
LDA & PG$_{\mathrm{nr}}$ (22\,d) & 0.961 \\
LDA & raw (24\,d) & 0.977 \\
MLP (64 hidden) & PG (26\,d) & 0.988 \\
\bottomrule
\end{tabular}
\end{table}

\subsubsection{Noise robustness}

Trained on clean features, the classifier inherits the razor-thin margins
of a deterministic forward model: two per cent measurement noise already
collapses the accuracy to 0.21. Noise augmentation (five noisy copies per
training sample at random noise level) restores robustness at a modest
clean-data cost (accuracy 0.76--0.80 without noise): the
noise-augmented PG classifier holds 0.74--0.78 at 2\,\% noise, 0.69 at
5\,\%, 0.59 at 10\,\%, 0.49 at 15\,\% and 0.41 at 20\,\%
(Fig.~\ref{fig:noisedomain}a). The comparison with \citep{sen2024} (about
80\,\% region detection at 20\,\% noise) must be made fairly: their noise
is added to the waveform and their Fourier/averaging chain denoises it
implicitly, and their task is nine coarse regions; the present task is
twelve fine-grained classes in which three stenosis grades of the same
segment must be separated, and the highest harmonics carry order-unity
relative error at 20\,\% noise. Per-class behaviour at 20\,\% noise
supports the physics: the ICA-terminus occlusions---whose signatures are
the large low-harmonic ratio excursions of Fig.~\ref{fig:spectra}b---keep
TPR 0.78--0.86, while the stenosis grades, whose carotid signatures are
sub-per-cent, fall to 0.04--0.09.

\subsubsection{Domain gap: anatomical variants}

\begin{figure}[!tb]
\centering
\includegraphics[width=\textwidth]{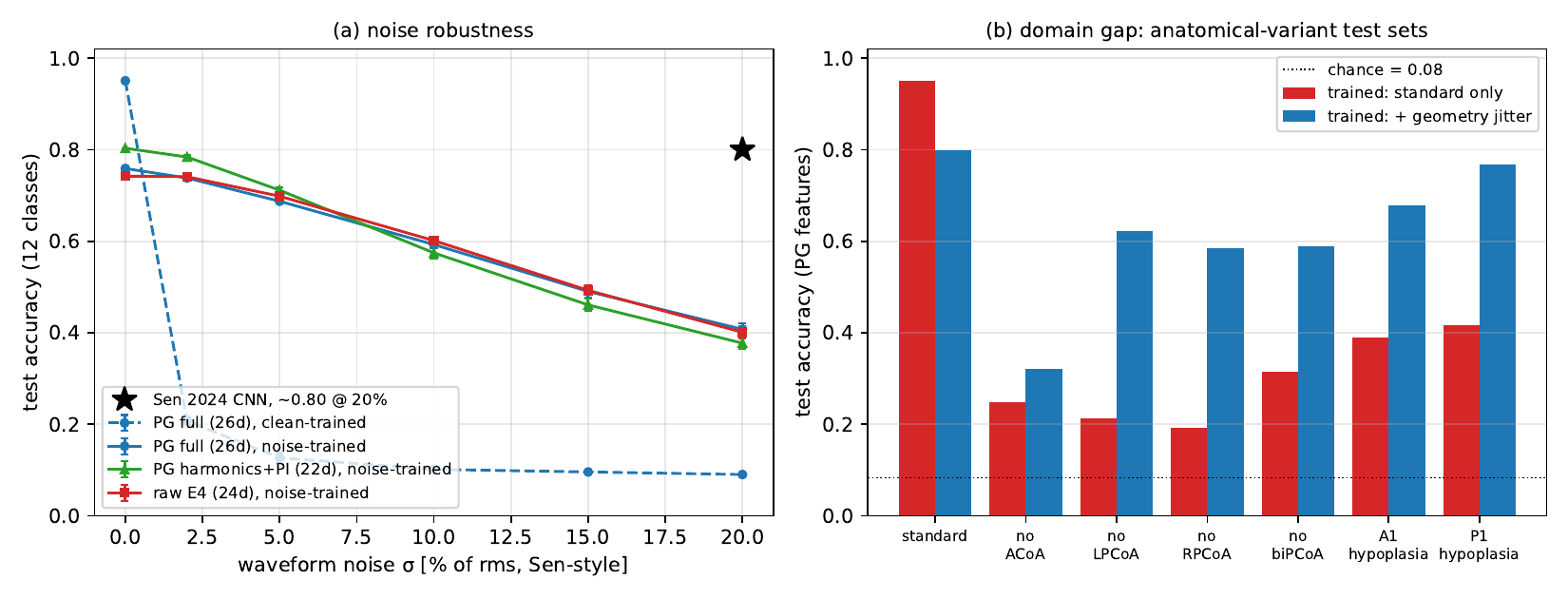}
\caption{Robustness of the physics-guided LDA classifier. (a) Accuracy
versus measurement-noise level for clean-trained and noise-augmented
classifiers (PG and raw features). (b) Domain gap and its mitigation:
accuracy of the standard-geometry-trained and geometry-augmented
classifiers on the standard geometry and on six anatomical-variant test
cohorts.}
\label{fig:noisedomain}
\end{figure}

A classifier trained on the standard geometry and tested on the six
anatomical-variant cohorts reproduces the failure mode documented for
waveform-driven CNNs on real asymmetric geometries \citep{sen2024}:
accuracy collapses to 0.19--0.42 depending on the variant
(Fig.~\ref{fig:noisedomain}b), worst for an absent right PCoA (0.19) or an
absent left PCoA (0.21)---configurations in which the posterior
collateral signature is structurally altered. The defence is equally
physical: augmenting the training cohort with the same anatomical variants
(cohort C) restores accuracy to 0.32--0.77 on the variant test sets while
keeping 0.80 on the standard geometry. The residual ranking is
informative: the absent-ACoA variant is the hardest (0.32), because the
anterior collateral route is removed entirely and the whole anterior
signature is rewritten, whereas P1 hypoplasia---a calibre change rather
than a topology change---is recovered to 0.77. Physics-guided features
outperform raw features on every variant after augmentation, consistent
with the healthy-model referencing removing part of the anatomy-dependent
baseline.

%=====================================================================
\section{Discussion}
\label{sec:discussion}

\subsection{What the scattering framework explains}

The starting observation of this study was empirical: data-driven
classifiers fed with carotid Doppler spectra can localise intracranial
occlusions, and their discriminative content concentrates at 2--12\,Hz
\citep{sen2024, argilaga2026}. The present framework supplies the missing
mechanism. Because the cerebral circulation sits deep in the acoustic
limit ($U/c\approx0.05$--$0.15$), the cardiac harmonics decouple, each
lesion scatters each harmonic independently with a strength set by
(\ref{eq:gammas}), and the round trip between the measurement station and
the lesion imprints the interference structure of (\ref{eq:ripple}). Three
consequences follow, all falsifiable and all borne out: the ripple spacing
$\Delta f=c_{\mathrm{eff}}/(2L_{\mathrm{eff}})$ for carotid geometry
(14.3--15.8\,Hz, side-stratified) lies above the discriminative band, so
the exploitable
localising information resides in harmonic amplitude ratios and ripple
depth, not in a resolved ripple; the band itself is selected by the
excitation spectrum, the $\sqrt{\omega}$ Womersley attenuation and the
frequency dependence of the scattering strength; and the band is
heart-rate-invariant only when expressed in harmonic orders. 

The
side-resolved asymmetry we predict for intracranial occlusions
(ipsilateral second harmonic attenuated to 0.92 for the
carotid-terminus occlusion, with the ipsilateral low-harmonic ratios
crossing unity near $n\approx6$) is a smaller-amplitude descendant of the
carotid velocity-ratio asymmetry (1.7--5.7) measured clinically for
proximal innominate occlusion \citep{grant2006}---smaller because the
reflection path is longer, exactly the distance scaling of
(\ref{eq:deltaf})---and the remotely generated waveform distortion of the
clinical tardus--parvus case \citep{park2019} is its qualitative mirror.
The synthetic spectrograms add the practical lesson that the envelope
extrema (PSV, EDV) are nearly silent to intracranial lesions at the
carotid station---a 90\,\% M1 stenosis moves the CCA PSV by half a per
cent---so any carotid-based localisation must exploit the pulsatility and
the harmonic content below the envelope, precisely the features the
physics-guided classifier uses. Where PSV does speak is at the lesion
itself: the computed throat velocities reproduce the bedside PSV-ratio
grading criterion \citep{nuffer2017}, so transcranial Doppler at the M1
and the network-inferred carotid signatures are complementary, not
competing, observables.

\iffalse
\subsection{The external benchmark, honestly read}

The Alastruey benchmark plays two roles. As a validation, it anchors the
network to an independent published model: healthy efferent flows agree
within 1--9\,\%, the collateral channels engaged by anatomical variants
within 11--16\,\%, and the communicating-artery calibre thresholds and
nominal collateral flows bracket the reference values. As a methodological
statement, it exposes how much of an occlusion-response prediction is
boundary-condition philosophy rather than model physics: with four
prescribed inlet flows the redistribution appears in the communicating
branches and the ICA/ECA split (frozen variant), while holding the total
headward inflow constant (compensated variant) mimics cardiac-output
redistribution---and the reference values of \citep{alastruey2007}, driven
by a single cardiac inflow, fall systematically between the two. The one
combination the present model gets wrong (absent right P1 with absent left
PCoA predicted as poor as the absent-A1 plus contralateral-ICA-occlusion
worst case) points at the two missing physical ingredients: the quadratic
pressure drop along communicating segments, which in the nonlinear model
penalises the high-flow collateral route, and a vertebral inflow that can
respond to the occlusion. Both are fixable within the framework---the
first by iterating the linearisation about the redistributed mean flow,
the second by closing the inlets with a heart model---and the benchmark
tells us exactly where they matter.
\fi

\subsection{Physics-guided features versus waveform-driven CNNs}

Read against the classification state of the art \citep{sen2024}, the
present results isolate what the physics buys. The headline accuracies are
comparable (95--99\,\% depending on features and classifier, against
$F_1\approx0.95$ for large-vessel-occlusion detection), but the sample
complexity differs by more than an order of magnitude---five samples per
class already suffice for 0.91 accuracy, against saturation around 7{,}000
waveforms---because the physics-guided features discard the
anatomy- and heart-rate-dependent baseline that a CNN must otherwise learn
from data. 

The same referencing buys transferability: after geometry
augmentation, PG features beat raw features on every anatomical variant.
The domain-gap experiment also reproduces, in controlled form, the failure
mode that \citep{sen2024} documented on real asymmetric geometries, and
shows both that the failure is structural (removing a collateral route
rewrites the signature, and no amount of standard-geometry training can
anticipate it) and that it is quantifiable and partially correctable
(augmentation restores 0.32--0.77 depending on the variant). The residual
weaknesses are equally informative: the healthy class and the mildest
stenosis grade are mutually confusable because their carotid signatures
are physically near-identical, and fine stenosis grading is the first
casualty of measurement noise---both are honest statements about the
information content of the carotid spectrum, not about the classifier.

\subsection{Limitations}

We should point out the, as an analytical model, the propose approach developed in this work contains 
four limitations. 

First, \emph{linearisation}:
the nonlinear solver comparison brackets the harmonic-magnitude error at
0.90--1.04, with the discarded quadratic Young--Tsai term visible in the
tightest stenoses; the signatures used here are ratios and are robust to
the side-independent systematics, but absolute-amplitude predictions carry
that error. 

Second, \emph{geometry}: the network is a single
population-mean anatomy with generic communicating-artery calibres; the
domain-gap experiments quantify how much that costs and how far
augmentation repairs it, but patient-specific anatomy from imaging is the
real answer. 

Third, \emph{synthetic circularity}: the cohort, the noise
model and the anatomical variants are all generated by the same model
class that defines the features, so the reported accuracies are
information-content statements, not clinical performance estimates; the
synthetic Doppler pipeline, although physically grounded, is a rendering
of the same one-dimensional physics and adds no independent tissue,
speckle or beam-formation physics of the kind full-physics ultrasound
simulation would. 

Fourth, \emph{boundary conditions}: prescribed-flow
inlets freeze the cardiac response, and the compensated variant is a
bracket, not a heart model. The decisive test is therefore prospective:
cycle-averaged carotid Doppler spectra from patients with
imaging-confirmed occlusion sites, against which the harmonic-ratio
signatures of Fig.~\ref{fig:spectra}b are directly comparable. We note
that the demonstrated tolerance of waveform classifiers to twenty per cent
measurement noise \citep{sen2024} places the predicted signature
amplitudes (harmonic-ratio excursions of eight to nine per cent,
sign-resolved between sides) within detectable range.

%=====================================================================
\section{Conclusions}
\label{sec:conclusions}

This study set out to answer a clinical question—whether the empirically observed occlusion-localising information in carotid Doppler spectra can be retrieved from a mechanics-based model—and our results show that it can. The spectral-decomposition scaffolding originally devised for simple-harmonic flow over topography transfers to the compliant circle-of-Willis network essentially intact, and it transforms the empirical observation ``low-frequency carotid Doppler spectra know where the occlusion is'' into a quantitative, interpretable scattering framework. In doing so, the model supplies the physical mechanism missing from purely data-driven classifiers: the lesion acts as a frequency-selective reflector, imprinting harmonic-ratio redistributions and a spectral ripple whose spacing encodes the occluder distance, while the empirically discriminative 2--12 Hz band is explained as the joint outcome of the cardiac excitation spectrum, the Womersley attenuation growing as \(\sqrt{\omega}\), and the frequency dependence of the scattering strength—a band that is more properly expressed in heart-rate-invariant harmonic orders.

Relative to the state of the art, the paper makes the following specific contributions.

\begin{enumerate}
    \item A linear frequency-domain scattering framework in which each cardiac harmonic propagates on a Womersley transmission line and is scattered independently by the lesion, yielding closed-form signatures: a ripple spacing \(\Delta f = c_{\mathrm{eff}}/(2L_{\mathrm{eff}})\) that encodes the occluder distance, harmonic-amplitude-ratio redistributions that encode site and severity, and a three-factor physical explanation of the empirically discriminative 2--12 Hz band, which is more properly expressed in heart-rate-invariant harmonic orders.
    \item Validation against an independent nonlinear one-dimensional solver across twelve lesion scenarios (mean flow divisions to three significant figures, harmonic magnitudes within 0.90--1.04, waveforms to 6.1\% in relative \(L_2\) after the area correction) and against three-dimensional stenosis computations at 75\% and 90\% area reduction, which confirm both the Young--Tsai loss model (cycle-mean drops 3.9 vs. 3.6 and 26 vs. 29 mmHg) and the bedside PSV-ratio grading criterion.
    \item An external benchmark that reproduces the healthy flow distribution, collateral recruitment and calibre thresholds of Alastruey et al.~[9], with the boundary-condition philosophy difference made explicit and bracketed by frozen-inflow and compensated-inflow variants.
    \item A physics-based synthetic spectral-Doppler pipeline that produces scanner-realistic spectrograms at seconds-per-scenario cost, linking the network model directly to the clinical observable and showing that the envelope extrema (PSV, EDV) are nearly silent to intracranial lesions at the carotid station, while the discriminative content lies in the pulsatility and harmonic structure below the envelope.
    \item A physics-guided feature set that localises twelve lesion classes at 95\% accuracy from five training samples per class—more than an order of magnitude below waveform-driven CNNs—with quantified noise and anatomical domain-gap behaviour and augmentation-based defences. The residual confusion is concentrated exactly where the physics predicts: between healthy and the mildest (50\%) M1 stenosis, whose carotid signature is sub-percent.
\end{enumerate}

The signatures derived here are directly comparable to cycle-averaged patient Doppler spectra, and the decisive next step is a prospective validation on real patient data with imaging-confirmed occlusion sites. The demonstrated tolerance of waveform classifiers to 20\% measurement noise places the predicted harmonic-ratio excursions (8--9\%, sign-resolved between sides) within detectable range, and the sample efficiency of the physics-guided features suggests that a mechanics-based pre-hospital localisation tool may be achievable without the need for large, expensive training cohorts.

\section*{CRediT authorship contribution statement}
\textbf{Xun Huang:} Conceptualization, Methodology, Software, Validation,
Formal analysis, Investigation, Data curation, Writing -- original draft,
Writing -- review \& editing, Visualization, Funding acquisition.

\section*{Declaration of competing interest}
The author declares that he has no known competing financial interests or
personal relationships that could have appeared to influence the work
reported in this paper.

\section*{Data availability}
The network configuration, all simulation and analysis codes, the
generated cohorts and the figure data will be made available in a public
repository upon acceptance; the external benchmark data are taken from the
published reference \citep{alastruey2007}.

\section*{Acknowledgements}
This work is partly supported by National Science Foundation of China (Grant No. 12432016 \& No. 12272007). 
The author has utilised Kimi to prepare the simulation running and sketching scripts.   

\bibliographystyle{elsarticle-num}
\bibliography{references}

\end{document}